\documentclass[acmsmall,screen,nonacm]{acmart}

\usepackage[utf8]{inputenc}
\usepackage[T1]{fontenc}
\usepackage{graphicx} 
\usepackage{listings}
\usepackage{microtype}
\usepackage{booktabs}
\usepackage{xspace}
\usepackage{tabto}
\usepackage{wrapfig}
\usepackage{comment}
\usepackage{proof}

\usepackage[disable]{todonotes}
\usepackage{subfig}
\usepackage[inline]{enumitem}
\usepackage{siunitx}
\usepackage{relsize} %
\usepackage{standalone}
\usepackage{fontawesome}
\usetikzlibrary{calc}

\definecolor{svblue}{HTML}{097cc3}   %
\definecolor{svparens}{HTML}{075d92} %
\definecolor{svdarkblue}{HTML}{053e61}

\definecolor{svgreen}{HTML}{2eb82e}  %
\definecolor{svdarkgreen}{HTML}{248f24}  %
\definecolor{svred}{HTML}{cc2900}    %
\definecolor{svdarkred}{HTML}{991f00}    %

\usepackage[capitalise,nosort,nameinlink]{cleveref}

\crefformat{footnote}{#2\footnotemark[#1]#3}
\crefname{section}{Sect.}{Sects.}
\Crefname{section}{Section}{Sections}
\crefname{algorithm}{Alg.}{Algs.}
\Crefname{algorithm}{Algorithm}{Algorithms}
\crefname{line}{line}{lines}
\crefalias{ALC@unique}{line}
\crefalias{ALC@line}{line}

\makeatletter
\def\blfootnote{\gdef\@thefnmark{}\@footnotetext}
\makeatother

\definecolor{darkgreen}{rgb}{0.0, 0.53, 0.0}
\definecolor{darkblue}{rgb}{0.21,0.21,0.50}
\definecolor{darkgrey}{rgb}{0.17,0.17,0.17}

\newcommand{\code}[1]{\texttt{#1}}

\newcommand{\alt}{\kern4.5pt\mathrel{\mid}\kern4.5pt}
\newcommand{\altspace}{\kern17.5pt}

\newcommand{\nonterminal}[1]{{\color{svdarkgreen} \langle \textit{#1} \rangle}}
\newcommand{\terminal}[1]{{\color{svdarkblue} \code{\bfseries #1}}}

\newcommand{\rplus}{{\color{darkgrey}^{+}}}
\newcommand{\rstar}{{\color{darkgrey}^{*}}}
\newcommand{\ropt}{{\color{darkgrey}^{?}}}

\newcommand{\LP}{\terminal{(}}
\newcommand{\RP}{\terminal{)}}

\newcommand{\QRY}{{\color{gray} \code{>}}}
\newcommand{\RES}{{\color{gray} \code{<}}}

\newcommand{\inlineheadingbf}[1]{\medskip\noindent{\bfseries #1.}}

\definecolor{basicLanguageColor}{HTML}{cca300}
\definecolor{procedureStatementsColor}{HTML}{cc6600}
\definecolor{unstructuredControlFlowColor}{HTML}{cc0000}
\definecolor{structuredControlFlowColor}{HTML}{86b300}
\definecolor{nondeterministicColor}{HTML}{cc0052}
\newcommand{\basicLanguageFragment}{{\color{basicLanguageColor} B}}
\newcommand{\procedureStatementsFragment}{{\color{procedureStatementsColor} P}}
\newcommand{\unstructuredControlFlowFragment}{{\color{unstructuredControlFlowColor} U}}
\newcommand{\structuredControlFlowFragment}{{\color{structuredControlFlowColor} S}}
\newcommand{\nondeterministicFragment}{{\color{nondeterministicColor} N}}

\definecolor{globallyColor}{HTML}{009973}
\definecolor{finallyColor}{HTML}{006bb3}
\newcommand{\safetyProperty}{{\color{globallyColor} G}}
\newcommand{\livenessProperty}{{\color{finallyColor} F}}

\newcommand{\reservedSymbolPrefix}{\code{\#}}

\newcommand\toolnamesize\smaller
\newcommand\tool[1]{{{\toolnamesize\scshape #1}\xspace}}
\newcommand\definetool  [2]{\newcommand  {#1}{\tool{#2}\xspace}}

\definetool{\cpachecker}  {CPAchecker}
\definetool{\benchexec}   {BenchExec}
\definetool{\pysvlib}{PySvLib}

\lstdefinelanguage{SVLIB}{
    basicstyle=\scriptsize\ttfamily,
    keywords=[1]{
        define-proc, define-procs-rec, declare-var,
        annotate-tag, select-trace,
        verify-call, get-witness, get-proof,
    },
    keywordstyle=[1]\color{svdarkblue}\bfseries,
    keywords=[2]{
        sequence, !, call, return,
        label, goto, if, while, break, continue,
        havoc, choice,
        assign,
        assume,
    },
    keywordstyle=[2]\color{svparens}\bfseries,
    keywords=[3]{:check-true, 
                :invariant,
                :not-recurring,
                :recurring,
                :requires, 
                :ensures, 
                :decreases, 
                :tag},
    keywordstyle=[3]\color{darkgreen}\bfseries,
    keywords=[4]{
        set-logic, declare-sort, define-sort,
        assert, check-sat, push, pop, exit,
        declare-const, define-fun, define-fun-rec,
        get-assertions, get-info, get-option, get-proof,
        get-unsat-core, get-value, get-assignment,
        set-info, set-option
    },
    keywordstyle=[4]\color{svdarkblue}\bfseries,
    keywords=[5]{Bool, Int, Real, String, RegLan, UFBV, UFNIA, UFLIA, UFNRA, Array, Set, Bag},
    keywordstyle=[5]\color{black}\bfseries,
    keywords=[6]{and, or, not, =>, =, <, <=, >, >=, +, -, *, div, mod, to_real, to_int, is_int,
        select, store, union, intersect, setminus, subset, member, cardinality,
        true, false
    },
    keywordstyle=[6]\color{black}\bfseries,
    keywords=[7]{1, 0, -1},
    keywordstyle=[7]\color{black}\bfseries,
    keywords=[8]{model, steps, incorrect-annotation, falsify-call, entry-proc, init-global-vars, 
                    init-proc-vars, leap, entry-call, choice, set-tag, using-annotation},
    keywordstyle=[8]\color{svdarkred}\bfseries,
    sensitive=false, %
    morecomment=[l]{;}, %
    alsoletter={-!:<=*+10},   %
    commentstyle=\color{gray}\ttfamily,
    moredelim=**[s][\color{black}](){}, %
} %

\lstdefinelanguage{MyC}{
    basicstyle=\scriptsize\ttfamily,
    keywords=[1]{
        while, return,
    },
    keywordstyle=[1]\color{svdarkblue}\bfseries,
    keywords=[2]{
        int,
    },
    keywordstyle=[2]\color{darkblue}\bfseries,
    sensitive=false, %
    morecomment=[l]{;}, %
    alsoletter={-!:<=*+10},   %
    commentstyle=\color{gray}\ttfamily,
    moredelim=**[s][\color{black}](){}, %
} %

\makeatletter
\pgfdeclareshape{document}{
  \inheritsavedanchors[from=rectangle]
  \inheritanchorborder[from=rectangle]
  \inheritanchor[from=rectangle]{center}
  \inheritanchor[from=rectangle]{north}
  \inheritanchor[from=rectangle]{south}
  \inheritanchor[from=rectangle]{west}
  \inheritanchor[from=rectangle]{east}
  \inheritanchor[from=rectangle]{north west}
  \inheritanchor[from=rectangle]{north east}
  \inheritanchor[from=rectangle]{south west}
  \inheritanchor[from=rectangle]{south east}
  \backgroundpath{
    \southwest \pgf@xa=\pgf@x \pgf@ya=\pgf@y
    \northeast \pgf@xb=\pgf@x \pgf@yb=\pgf@y
    \pgf@xc=\pgf@xb \advance\pgf@xc by-6pt
    \pgf@yc=\pgf@yb \advance\pgf@yc by-6pt
    \pgfpathmoveto{\pgfpoint{\pgf@xa}{\pgf@ya}}
    \pgfpathlineto{\pgfpoint{\pgf@xa}{\pgf@yb}}
    \pgfpathlineto{\pgfpoint{\pgf@xc}{\pgf@yb}}
    \pgfpathlineto{\pgfpoint{\pgf@xb}{\pgf@yc}}
    \pgfpathlineto{\pgfpoint{\pgf@xb}{\pgf@ya}}
    \pgfpathclose
    \pgfpathmoveto{\pgfpoint{\pgf@xc}{\pgf@yb}}
    \pgfpathlineto{\pgfpoint{\pgf@xc}{\pgf@yc}}
    \pgfpathlineto{\pgfpoint{\pgf@xb}{\pgf@yc}}
    \pgfpathlineto{\pgfpoint{\pgf@xc}{\pgf@yc}}
  }
}
\makeatother

\definecolor{tolDarkBlue}{HTML}{332288}
\definecolor{tolDarkGreen}{HTML}{117733}
\definecolor{tolTeal}{HTML}{44aa99}
\definecolor{tolLightBlue}{HTML}{88ccee}
\definecolor{tolYellow}{HTML}{ddcc77}
\definecolor{tolRed}{HTML}{cc6677}
\definecolor{tolViolet}{HTML}{aa4499}
\definecolor{tolBurgundy}{HTML}{882255}
\definecolor{tolPastelTeal}{HTML}{d0eae6}
\definecolor{tolPastelGreen}{HTML}{c4ddcc}
\definecolor{tolPastelBlue}{HTML}{e1f2fb}
\definecolor{tolPastelRed}{HTML}{f2d9dd}
\definecolor{tolPastelYellow}{HTML}{f6f2dd}
\definecolor{tolPastelViolet}{HTML}{ead0e6}
\definecolor{tolPastelBurgundy}{HTML}{e1c8d5}
\definecolor{colorAlmostWhite}{HTML}{f5f7f7}
\definecolor{pastelOrange}{HTML}{F39949}
\definecolor{pastelGreen}{HTML}{B9CC67}

\tikzset{
    st/.style={
        font=\ttfamily,
        shape=rectangle,
        rounded corners=.5em,
        fill=gray!40,
        inner xsep=.3em,
        inner ysep=0em,
        text height=2ex,
        text depth=.6ex
    },
    actor edge/.style={
        draw=black,
        ->,
        shorten >=2pt,
        shorten <=2pt,
        >=latex, 
    },
    bi actor edge/.style={
        draw=black,
        <->,
        shorten >=2pt,
        shorten <=2pt,
        >=latex, 
    },
    obj/.style={
        draw,
        align=center,
        minimum height=1.1cm,
        font=\sffamily,
    },
    actor/.style={
        obj,
        fill=colorAlmostWhite,
    },
    artifact/.style={
        obj,
        draw=black,  %
        minimum height=.5cm,
        minimum width=.5cm,
        inner sep=2pt,
        fill=colorAlmostWhite,
        document,
        rounded corners=1pt,
    },
    artifact label/.style={
        font=\sffamily\scriptsize,
        inner sep=0pt,
        color=black,
    },
    program/.style={
        artifact,
        fill=tolPastelBlue,
    },
    spec/.style={
        artifact,
        fill=tolPastelYellow,
    },
    config/.style={
        artifact,
        fill=tolPastelTeal,
    },
    verdict/.style={
        artifact,
        left color=tolPastelGreen,
        right color=tolPastelRed,
    },
    report/.style={
        artifact,
        shading=axis, shading angle=90,
        left color=gray,
        right color=white,
    },
    witness/.style={
        artifact,
        fill=tolPastelBurgundy,
    },
    testcase/.style={
        artifact,
        fill=tolPastelRed,
    },
    smt/.style={
        artifact,
        fill=tolLightBlue,
    },
    k2/.style={
        artifact,
        fill=pastelGreen,
    },
    moxi/.style={
        artifact,
        fill=tolPastelViolet,
    },
    why3/.style={
        artifact,
        fill=tolTeal,
    },
    viper/.style={
        artifact,
        fill=tolBurgundy,
    },
}

\newcommand{\webpageurl}{\url{https://gitlab.com/sosy-lab/benchmarking/sv-lib}\xspace}

\newcommand{\TODO}[1]{}
\newcommand{\DISCUSS}[1]{}
\newcommand{\RESOLVED}[1]{}
\newcommand{\OBSOLETE}[1]{}
\newcommand{\WONTFIX}[1]{}
\newcommand{\Language}{\mbox{SV-LIB}\xspace}

\DisableLigatures{family=tt*}

\begin{document}

\title[SV-LIB 1.0: A Standard Exchange Format for Software-Verification Tasks]{%
    SV-LIB 1.0: A Standard Exchange Format for\\
    Software-Verification Tasks
}

\author{Dirk Beyer}
\orcid{0000-0003-4832-7662}
\affiliation{%
  \department{Institute for Informatics}
  \institution{LMU Munich}
  \city{Munich}
  \country{Germany}
}
\email{dirk.beyer@sosy.ifi.lmu.de}

\author{Gidon Ernst}
\orcid{0000-0002-3289-5764}
\affiliation{%
  \department{Institute for Informatics}
  \institution{LMU Munich}
  \city{Munich}
  \country{Germany}
}
\email{gidon.ernst@sosy.ifi.lmu.de}

\author{Martin Jonáš}
\orcid{0000-0003-4703-0795}
\affiliation{%
  \department{Faculty of Informatics}
  \institution{Masaryk University}
  \city{Brno}
  \country{Czech Republic}
}
\email{martin.jonas@mail.muni.cz}

\author{Marian Lingsch-Rosenfeld}
\orcid{0000-0002-8172-3184}
\affiliation{%
  \department{Institute for Informatics}
  \institution{LMU Munich}
  \city{Munich}
  \country{Germany}
}
\email{marian.lingsch-rosenfeld@sosy.ifi.lmu.de}

\begin{abstract}
    In the past two decades, significant research and development effort went into
the development of verification tools for individual languages, such as~C, C++,
and Java. Many of the used verification approaches are in fact language-agnostic
and it would be beneficial for the technology transfer to allow for using the
implementations also for other programming and modeling languages. To address
the problem, we propose \Language, an exchange format and intermediate language
for software-verification tasks, including
programs, specifications, and verification witnesses.
\Language~is based on well-known concepts from imperative programming languages
and uses SMT-LIB to represent expressions and sorts used in the program. This
makes it easy to parse and to build into existing infrastructure, since many
verification tools are based on SMT solvers already. Furthermore,
\Language~defines a witness format for both correct and incorrect \Language
programs, together with means for specifying witness-validation tasks. This
makes it possible both to implement independent witness validators and to reuse
some verifiers also as validators for witnesses.
This paper presents version 1.0 of the \Language~format, including its
design goals, the syntax, and informal semantics. Formal semantics and further
extensions to concurrency are planned for future versions.

\end{abstract}


\maketitle

\vspace{3.0cm}

\begin{center}
     {\LARGE Version 1.0} \qquad\qquad
     \raisebox{-3.4mm}{\includegraphics[width=3.3cm]{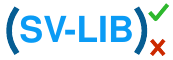}} \qquad\qquad 
     {\LARGE 2025-11-26} \\[0.5cm]
     {\LARGE \webpageurl}
\end{center}

\newpage
{
\footnotesize
\tableofcontents
}
\section{Introduction}
\label{sec:introduction}
Software verification of a given input program~$p$ and input
specification~$\varphi$ is the problem to determine whether the
statement~$p\models\varphi$ holds, i.e., whether $p$~satisfies all properties
in~$\varphi$. The specification can describe safety and liveness properties,
such as absence of runtime errors, functional correctness, and termination. If
the verifier finds a violation to the specification, it provides a violation
witness~\cite{Witnesses}, which testifies the specification violation by describing a
counterexample.
If the verifier can prove that the specification holds, then it provides a
correctness witness~\cite{CorrectnessWitnesses}, which testifies that the specification holds. The precise
form of the correctness witness depends on the property, and can for example
consist of loop invariants for safety and of ranking functions for proving
termination.
Witness validation
for a given input program~$p$, input specification~$\varphi$, and witness~$w$
is the problem of determining whether the statement~$p\models_w\varphi$ holds, that is,
whether the information in~$w$ is correct and can be used for the proof of
$p\models\varphi$ or $p\not\models\varphi$.

A recent Dagstuhl seminar~\cite{Dagstuhl25-InfoEx} noted that,
while there are many state-of-the-art tools for software verification available,
the coverage of the various programming languages is very diverse.
For example, there are many verifiers for the languages~C and~Java,
but not so many verifiers for the languages~C++ and~Rust.
A recent overview~\cite{SoftwareModelChecking20Years} noted that
there has been immense progress in research and technology in the past two decades,
leading from a lack of verification tools to an abundance of them:
the overview counts 76~automatic verifiers for the languages~C and~Java.

Most verification approaches are not specific to a particular programming language,
but of general nature.
For example, bounded model checking~\cite{BMC}, k-induction~\cite{kInduction},
property-driven reachability~\cite{IC3}, interpolation-based model checking (IMC)~\cite{McMillanCraig},
and others~\cite{VizelFMCAD09,ForwardBackwardReachability} are implemented in many
verification tools (see~\cite{SVCOMP25}, Table~8).
Those approaches were usually first invented for hardware verification and
later adopted to software verification (e.g., \cite{IMC-JAR,DAR-transferability}).
It is worth noting that this can take 20 years, as for the original IMC algorithm~\cite{IMC-JAR}.

The verification approaches and other new technology can be transferred much
faster to other languages via \emph{transformation} of the input program or
model (for example, from~Btor2 to~C~\cite{BTOR2C} and
back~\cite{Kratos2,CPV-SVCOMP24}). It is not necessary to re-implement each
verification approach, but only a transformation needs to be implemented once and
then existing tools can be applied. Similarly, many approaches are readily
available for the language~C but not for many other languages: An
intermediate language for programs could help in this situation.

Furthermore, there is an acute lack of information exchange between verifiers form different
communities working on software verification. In particular, it is difficult to
exchange information between deductive verification tools
(e.g.,~\cite{Dafny,BoogieVerifier,Viper}) and automatic software verification
tools
(e.g.,~\cite{CPAchecker-3.0-tutorial,CBMC,UAUTOMIZER-SVCOMP24}),
although preliminary studies~\cite{AutoActive,VerCors-CPA} showed that it can be helpful.
This is partially due to the different input languages used and the differing
goals and semantics of the tools. A common intermediate language for programs
with a well-defined semantics could help here as well.

\begin{figure}[t]
  \centering
	\input{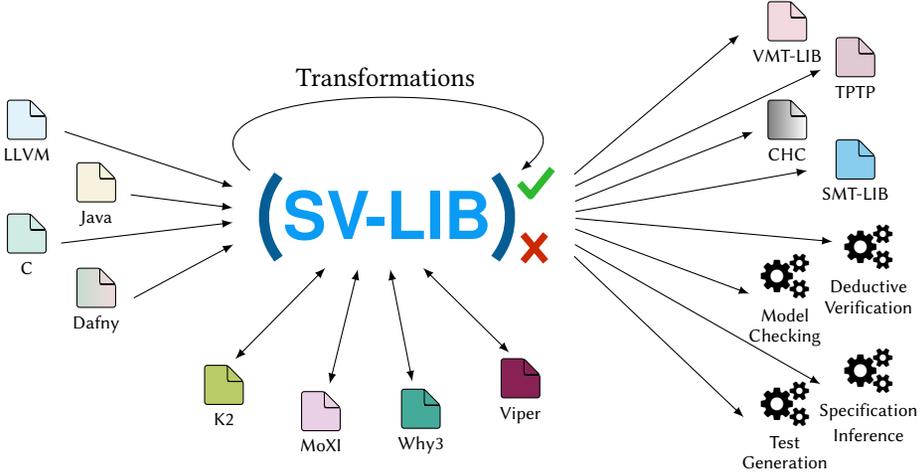}
  \vspace{-2mm}
    \caption{
      Possible transformations and use-cases for \Language~as an intermediate language
    }
    \label{fig:transformation}
  \vspace{0mm}
\end{figure}

An intermediate language can be used also to transform programs from an imperative
programming language~A to an imperative programming language~B
(A~$\to$~\Language and \Language~$\to$~B),
for example with the goal to use a verifier for~B to verify programs written in~A.
This needs only $\mathcal{O}(n)$~transformations for~$n$
source and target languages. 
Using direct transformations between~$n$ end-user
programming languages, we would need to develop $\mathcal{O}(n^2)$~transformations.
This is illustrated in \cref{fig:transformation}.
In particular, through transformation, the different communities
(hardware vs. software, deductive vs. automatic, models vs. programming
languages) grow together. This can be observed by the extension of benchmark
sets for competitions: For example, there is a benchmark set
originating from the hardware-verification competition HWMC (which
focusses on Btor2 and Aiger) that was contributed into the software-verification
competition SV-COMP (which focussed on~C and Java). 
In a similar vein, we translated C programs into \Language programs 
(\href{https://gitlab.com/sosy-lab/benchmarking/sv-benchmarks/-/tree/main/sv-lib}{sv-benchmarks/.../sv-lib}).
Another example for inter-community exchange is using the exchange format for
verification witnesses~\cite{VerificationWitnesses-2.0} for cooperation
between automatic and interactive verifiers~\cite{VerCors-CPA}.

Multiple intermediate formats have been developed over the past years, and are
in widespread use, prominently Boogie~\cite{Boogie}, Why3~\cite{Why3}, and Viper~\cite{Viper}.
In contrast to \Language, these are not directly based on SMT-LIB,
making their parsing and support more difficult for tools,
and they incorporate some higher-level features such as modules (Why3),
expressive types (Why3, Boogie), or impose a heap model (Viper),
which offers convenience in some situations but hinders adoption for other use-cases.
The intermediate languages 
VMT~\cite{VMT} and MoXI~\cite{MoXI-Language,MOXICHECKER,MoXI-Tool-Suite}
are based on SMT-LIB, but they are meant for models and represent transition systems,
and do not support control flow. 
Furthermore, their support for witnesses is limited,
and not expressible in the input format itself.
Most closely related to \Language is the language K2~\cite{Kratos2},
which is an intermediate language for imperative programs (supporting \texttt{if} and \texttt{goto})
and served as the major inspiration for developing \Language.
\Language extends the spirit of K2 by adding
commands to express witnesses 
(including invariants, statement contracts, and ranking functions for correctness witnesses,
as well as concrete traces for violation witnesses), and 
structured control flow (\texttt{while}) directly to to the language.
In contrast to K2, \Language re-uses the SMT-LIB standard~\cite{SMTLIB27} verbatim,
supporting its type and term language as well as many of the SMT-LIB top-level commands.
In contrast to MoXI, we do not make a-priori restrictions on the allowed features, inheriting SMT-LIB's full expressiveness to cover use-cases in deductive verification.
As an added benefit, we largely avoid the effort to document and maintain that aspect of the language.

Another popular choice for an intermediate solver-backend language
is the lower-level format of Constrained Horn Clauses
(CHC)~\cite{ConstraintLogicPrograms,ConstraintLogicProgrammingSurvey,CHC}.
CHCs are standardized as a subset of the SMT-LIB language, which means
they offer a language- and tool-agnostic way to decouple front-ends of
verification tools from back-end solvers. However, the encoding of
a program-verification task into CHCs already makes some choices about the verification
approach to be used and discards higher-level structure of the original task
(e.g., invariants vs. loop contracts~\cite{LoopVerificationContractsInvariants},
large blocks vs. small blocks~\cite{LBE,ABE}).
Moreover, despite some efforts, the format of models and proofs for CHC problems
is not standardized across verifiers and the information these artifacts convey is
difficult to relate back to the original tasks.

Not only are intermediate languages useful for verification,
they are also a popular choice for working with and 
compiling programming languages.
For example, LLVM~IR~\cite{LLVM-CGO04} is a widely-used
intermediate representation for optimizing compilers and program analysis tools.
Other frameworks, such as Java Bytecode and WebAssembly,
serve as intermediate representations for their respective ecosystems.
In such environments, having a flexible intermediate verification language
that covers a large span of technologies may help with the challenges of
multilingual verification tasks~\cite{ChallengesMultilingual}.

We developed the intermediate language for verification tasks \Language~to
address these problems. Verification approaches that are implemented for
\Language~programs can be used for various imperative languages by transforming
the verification tasks to \Language~and then applying any \Language~verifier.
Like CHC, we build on the SMT-LIB language and logic for the
representation of terms of the language, which covers a rich spectrum from
implementation-centric properties about arithmetic and arrays to behavioral
specifications that make use of data abstractions. In contrast to CHC, we aim
to \emph{capture program-verification tasks declaratively} and to \emph{enable
  justification of verdicts via a fully-specified witness format in the same
  format as the original verification task}. Thus, the format extends SMT-LIB
with a syntax for imperative program statements and commands for the definition
of procedures and correctness claims.

\section{Language Design}

This document introduces version~1.0 of \Language,
extending \mbox{SMT-LIB} version~2.7~\cite{SMTLIB27} with
constructs usually found in imperative programming languages,
like (recursive) procedures, (global) variables, and (un-)structured control flow.
In particular, extending SMT-LIB allows us to
reuse most of its infrastructure
like sorts, terms, functions, and some commands.
To differentiate between SMT-LIB files and \Language~files,
we propose to use the file ending \code{.svlib} for \Language~files,
and its corresponding witnesses.
In addition we recommend setting the option \terminal{:format-version}
to the version of the language that is used in the program,
such that it is possible to easily analyze the format version 
that the program corresponds to.

\subsection{Community Call for a New Intermediate Language}
The community proclaimed the need to exchange verification tasks and
proposed to create a new intermediate language~(IL) that
is easy to support and process by tools~\cite{Dagstuhl25-InfoEx}.
In the following, we list goals and requirements
for the new language, which were discussed in a working group at Dagstuhl~\cite{Dagstuhl25-InfoEx}.
The new language should:
\begin{enumerate}[label=(\alph*)]
  \item support the exchange of complete verification tasks, consisting of
    (1)~programs,
    (2)~specifications (safety and liveness), and
    (3)~witnesses (for counterexamples and proofs),
  \item be based on a human-readable text format (instead of a binary format)
    and use a well-known and easy to parse base syntax (e.g., Lisp-based, as used by SMT-LIB),
  \item have a formal semantics without undefined behavior
    (except for cases akin to SMT-LIB where the notion of under-specification
     is logically well-understood (such as that division by zero yields some unknown natural number)),
    such that comparative evaluations of verification algorithms do not depend on interpretations,
    and the same should hold for witness validation,
  \item support interoperability, such that frontends and
    backends are exchangable,
  \item express verification and validation tasks in the same language,
  \item allow that witness validation is compilable to simple first-order checks (i.e., SMT queries)
    without further inference, for all properties,
    that is, witnesses should be informative enough to allow validation with SMT Solvers,
  \item be independent from the higher-level input
    language (such as C and Java) and independent from verification
    technology used in the backend,
  \item have first-class support for widely-used concepts like global variables and procedures,
  \item offer a way to keep the specification separate from the program,
  \item retain more structure than MoXI, VMT, and CHC (control flow, procedures, sequential composition, ...)
    to express tasks before encoding to verification conditions by a particular approach,
  \item be based on well-understood programming concepts and consolidate existing languages
    in that spectrum (e.g., Boogie, Why3, Viper), and
  \item be usable for use-cases in both automatic and deductive verification.
\end{enumerate}

Furthermore, the community requested that the organizers of SV-COMP~\cite{SVCOMP25}
include a track on \Language in the 
Competition on Software Verification 2026\footnote{\url{https://sv-comp.sosy-lab.org/2026/demo.php}}.

\subsection{Design Principles for the New Intermediate Language \Language}

For ease of use, we build on a well-established and well-understood standard (SMT-LIB)
to encode the logical language of expressions.
We extend it with constructs native to an imperative language like
(recursive) procedures, (global) variables, and (un-)structured control flow,
remaining as close as possible to the existing syntax.
To address a wide range of approaches, the language is built around
a small core with several fragments, between which translations are possible.
For example, structured and unstructured control-flow fragments.

In the same manner as SMT solvers keep an interactive connection
with the solver, \Language~also keeps an interactive connection
with the verifier.
We describe the interaction from the point of view of the client of the verifier,
where ``$\RES$'' is a response of the verifier
and ``$\QRY$'' is a follow-up query from the client.
\vspace{-2mm}
\begin{align*}
\QRY~ & \LP \terminal{verify-call}~
            \nonterminal{symbol}~
            \LP \nonterminal{term}\rstar \RP
        \RP \notag \\
\RES~ & \terminal{correct} \qquad \text{or} \qquad \terminal{incorrect} \\[4pt]
\QRY~ & \LP \terminal{get-witness} \RP \\
\RES~ & \nonterminal{witness}
\end{align*}
\vspace{-3mm}

Verifiers should be able to read a list of commands from
standard input and write the outputs to standard output.
Similarly to SMT-LIB, verifiers can be controlled by setting options in the commands.
All verifiers are encouraged to support
the options \terminal{:regular-output-channel} and \terminal{:diagnostic-output-channel}.

Moreover, to make use of the compositional nature of the
specification mechanism in terms of \terminal{:annotate-tag} etc.,
verifiers should also accept verification tasks given as a list of files
on the command line, with the interpretation that their concatenation
in the given order jointly defines a single verification script.
For example, this allows users to use several specifications for the same program,
alternatively or conjunctively.
Also, a benchmark definition might be prepended
with use-case specific options (e.g., experimental evaluations might define that the witness
is printed to a specific file).

\subsection{Tool Standardization}

Apart from having a standard language to express verification tasks,
it has been recognized as beneficial to standardize the way tools
interact with each other, since this enables 
cooperative verification
and makes it easier for users to try out how other 
tools perform for the same task~\cite{FMWECK,CooperativeVerification}.

In contrast to SMT-LIB, where the interaction is based on commands
sent via standard input and output,
we propose to allow this and additionally standardize a set of 
command-line options which tools need to support. 
This approach is motivated by security concerns, since some
options may result in security concerns if set via the input file,
for example, if they influence the file system access of the tool.
On the other hand, it is useful to be able to set
some options via the input file to be able to document
those settings together with the verification task.

In order to achieve tool standardization,
\Language~defines the following set of command-line flags,
which verification tools should support.
Usually when a command-line flag is not security-relevant,
it can also be set via an option inside the input file.
Furthemore, the command-line flags have precedence over the options
set inside the input file overriding whatever was set in the program.
Currently tools supporting \Language~should support:

\begin{enumerate}[label=(\alph*)]
    \item ``\texttt{--produce-witnesses}'' which enables witness production, 
        it can also be set via the option
        \texttt{:produce-witnesses} inside the input file,

    \item ``\texttt{--witness-output-channel <channel>}'' which specifies
        where to output the witness,
        where \texttt{<channel>} is either
        \texttt{stdout}, \texttt{stderr}, or a POSIX compliant file name,
        If not specified it defaults to \texttt{stdout}.
\end{enumerate}
\vspace{-2mm}

\section{Full Examples}
\label{sec:motivation}

Before showing the detailed definition
of the syntax in~\cref{sec:commands,sec:statements,sec:properties,sec:witnesses,sec:well-formed-programs}
including the informal semantics of \Language,
we provide some full examples (based on the algorithm \texttt{add} 
from \cref{fig:correct-program-c}, 
whose variables are encoded as unbounded 
integers in the \Language examples for simplicity), showcasing
different scenarios and
objects (commands, properties, witnesses)
that users will encounter when working with \Language.

\begin{figure}[b]
    \vspace{-5mm}
                \lstinputlisting[
                        language=MyC,
                        numbers=none,
                        stepnumber=1,
                        firstnumber=1,
                        numberfirstline=true,
                        xleftmargin=55mm]
                {
                    examples/simple-correct.c
                }
    \vspace{-5mm}
    \caption{Procedure \texttt{add} written in C}
    \label{fig:correct-program-c}
\end{figure}

We first show an example of how the output
of the verifier looks when
verifying a correct program simultaneously 
for safety and liveness properties
in~\cref{sec:verifying-correct-program}.
Notable for this case, the verifier has to
produce a witness for both properties at the same time,
instead of a witness for safety and liveness
separately.
Afterwards, we extend this example to show an
application to incremental verification
in~\cref{sec:incremental-solver-interaction}.
In this case the verifier can choose to maintain an
internal state across multiple \terminal{verify-call} commands,
in order to be more efficient.
This is not required, and the verifier can re-verify the
full program if it cannot perform stateful verification.

Finally we show multiple examples of how the output
of the verifier looks like when
verifying a program with an incorrect safety property
(\cref{sec:verifying-incorrect-safety-property}),
liveness property
(\cref{sec:verifying-incorrect-liveness-property}),
and invariants
(\cref{sec:verifying-incorrect-invariant}).
These examples illustrate how a violation witness
looks for different types of properties.
In particular, the
violation witness for invariants
in~\cref{sec:verifying-incorrect-invariant} illustrates
how modular correctness annotations are refuted,
which may make reference to paths which do not exist in the program
but which arise only in the overapproximation
expressed by the inductiveness of invariants and contracts
and the well-foundedness conditions of ranking functions.

After showing the use-cases for verification, we briefly
show in~\cref{sec:witness-validation-example}
how a witness and program can be linked together
in order to generate a validation task.
This is important, since both validation tasks and
verification tasks are \Language~scripts,
and therefore, one needs to be able to combine
a verification task with a witness into a
new \Language~script.

\subsection{Verifying a Correct Program}
\label{sec:verifying-correct-program}

\begin{figure}[t]
    \hspace{1.0mm}
    \subfloat[Correct \Language~script]{
      \parbox{0.52\textwidth}{
        \lstinputlisting[
                    language=SVLIB, 
                    numbers=left,
                    stepnumber=1,
                    firstnumber=1,
                    numberfirstline=true,
                    numbersep=2pt,
                    numberstyle=\tiny,
                    escapechar=@]
        {
            examples/simple-correct.svlib
        }
        \label{fig:verifying-correct-program}
      }
    }%
      \subfloat[Possible witness for~\cref{fig:verifying-correct-program}]{
        \raisebox{-32.7mm}{
            \parbox{0.40\textwidth}{
                \lstinputlisting[
                        language=SVLIB,
                        numbers=none,
                        stepnumber=1,
                        firstnumber=1,
                        numberfirstline=true,
                        escapechar=@]
                {
                    examples/simple-correct-witness.svlib
                }
            }
        }
        \label{fig:correct-program-witness}
      }
    \vspace{-1mm}
    \caption{Correct \Language program that fulfills all
        its specifications (left) and a possible witness
        for it (right); scripts available at 
        the~\href{https://gitlab.com/sosy-lab/benchmarking/sv-lib}{\Language} repo
        as \href{https://gitlab.com/sosy-lab/benchmarking/sv-lib/-/blob/format-1.0/examples/core-verification/loop-add.svlib}{verification}
        and \href{https://gitlab.com/sosy-lab/benchmarking/sv-lib/-/blob/format-1.0/examples/core-validation/loop-add.svlib}{validation} tasks}
    \vspace{-4mm}
\end{figure}

Consider the correct program
in~\cref{fig:verifying-correct-program},
which uses the \terminal{LIA} SMT-LIB logic.
It defines in~\cref{line:correct-program-proc-add}
the procedure \terminal{add}
with two input constants~\terminal{x0} and~\terminal{y0},
an output variable~\terminal{x}, and
a local variable~\terminal{y}.
The procedure adds the two numbers \terminal{x0} and \terminal{y0}
using the local variable~\terminal{y},
and variable~\terminal{x} returns the result.
This program shows how \Language can separate the program
from its specification, since the specifications
are given by the \terminal{:annotate-tag}
commands, which refer to the \terminal{:tag} annotations 
and are therefore not intertwined
with the program itself.
This also allows the reuse of the same specifications at
multiple program locations.
Furthermore, the specification can also be added directly
by inlining them at the relevant locations, if desired.
The contract for the body of the 
procedure \terminal{add}
is given by the \terminal{:annotate-tag}
command in~\cref{line:correct-program-specification}.
Furthermore,~\cref{line:correct-program-loop-specification}
specifies that the loop 
in~\cref{line:correct-program-loop} terminates.
Finally, the \terminal{verify-call} command
in~\cref{line:correct-program-verification-call}
asks the verifier to prove that the procedure \terminal{add}
fulfills its specification for two constant inputs
\terminal{x1} and \terminal{y1}, which are defined
using the SMT-LIB command \terminal{declare-const}.

\Cref{fig:correct-program-witness}
shows an example witness for the program
in~\cref{fig:verifying-correct-program}.
It has the same structure as a \Language~script
itself, since it uses the same commands.
Having a program and its witness in the same format
allows tool developers to focus on a single format,
which can be used for both input and output,
and similarly for both verification and validation.

The goal of a correctness witness
is to simplify the validation of correctness,
when compared to verification from scratch.
Therefore, the witness provides
the necessary invariants, contracts,
and ranking functions,
to reduce the correctness proof
into SMT queries.

\subsection{Incremental Solver Interaction}
\label{sec:incremental-solver-interaction}

\begin{figure}[t]
    \hspace{-1.6mm}
    \subfloat[First part]{
      \parbox{0.52\textwidth}{
        \lstinputlisting[language=SVLIB, 
                     numbers=left,
                     firstnumber=1,
                     numberfirstline=true,
                     escapechar=@,
                     numbersep=2pt,
                     numberstyle=\tiny]
        {
            examples/simple-incremental-part1.svlib
        }
      }
    }%
    \subfloat[Second part]{
      \parbox{0.40\textwidth}{
        \raisebox{-55.8mm}{
            \lstinputlisting[language=SVLIB, 
                        numbers=left,
                        firstnumber=29,
                        numberfirstline=true,
                        escapechar=@,
                        numbersep=2pt,
                        numberstyle=\tiny]
            {
                examples/simple-incremental-part2.svlib
            }
        }
      }
    }
    \vspace{-1mm}
    \caption{Two parts of an \Language~script showing incremental solver interaction; the
        solver may choose to maintain a state across multiple
        \terminal{verify-call} commands to be more efficient;
        script available at
        the~\href{https://gitlab.com/sosy-lab/benchmarking/sv-lib}{\Language} repo
        as \href{https://gitlab.com/sosy-lab/benchmarking/sv-lib/-/blob/format-1.0/examples/core-verification/loop-add-incremental.svlib}{verification task}}
    \label{fig:incremental-solver-interaction}
    \vspace{-4mm}
\end{figure}

\Language~verifiers maintain a dialogue with the user, similar to SMT solvers.
The user can send multiple commands to the verifier,
and the verifier responds to each command.
This interaction benefits from the solver maintaining
its state across commands, allowing for more efficient
and context-aware responses.
Nonetheless, even though it may be useful to retain the state between verification calls,
this is left to the discretion of the solver, which has the freedom to decide if
it wants to keep parts of the state or recompute everything
from scratch.

\Cref{fig:incremental-solver-interaction} shows
an example in which a solver could benefit from incremental
verification. 
The first \terminal{verify-call} command
at~\cref{line:incremental-solving-first-verification}
proves the procedure \terminal{add} correct.
Therefore the contract of \terminal{add}
can be used to prove the correctness
of the procedure \terminal{mul} in the second
\terminal{verify-call} command
at~\cref{line:incremental-solving-second-verification}.
While a solver can reuse invariants, contracts counterexamples which it has
previously found, this may not always be useful.
Sometimes, a previous abstraction may not be strong
enough to prove the new property correct.
For example, the contract \terminal{:ensures} \terminal{true}
for the procedure \terminal{add}
may not be sufficient to prove the correctness
of the procedure \terminal{mul}.

The reuse of the internal state of the verifier is not only 
possible for finding invariants and contracts,
but also for finding counterexamples. For example one 
could verify from the start of the \terminal{verify-call}
command untill a state implying a previously found
counterexample is reached or reuse previously found
abstractions to speed up the search for a 
new counterexample~\cite{PrecisionReuse}.
But this should be done with care, since the previously
found abstractions may not be useful in the new context.

\subsection{Verifying an Incorrect Program for a Safety Property}
\label{sec:verifying-incorrect-safety-property}

\begin{figure}[t]
    \vspace{0mm}
    \hspace{2mm}
    \subfloat[\Language~script violating a safety property]{
      \parbox{0.52\textwidth}{
        \lstinputlisting[
                        language=SVLIB, 
                        numbers=left,
                        stepnumber=1,
                        firstnumber=1,
                        numberfirstline=true,
                        numbersep=2pt,
                        numberstyle=\tiny,
                        escapechar=@]
        {
            examples/simple-incorrect-safety.svlib
        }
    }
        \label{fig:verifying-incorrect-safety-property}
    }%
    \subfloat[Example witness for~\cref{fig:verifying-incorrect-safety-property}]{
        \raisebox{-2.04cm}{
            \parbox{0.40\textwidth}{
                \lstinputlisting[
                            language=SVLIB, 
                            numbers=none,
                            stepnumber=2,
                            firstnumber=1,
                            numberfirstline=true,
                            escapechar=@]
                {
                    examples/simple-incorrect-safety-witness.svlib
                }
            }
        }
        \label{fig:incorrect-safety-witness}
    }
    \vspace{-1mm}
    \caption{Incorrect \Language program violating a safety
        property (left) and a possible witness for it (right),
        which provides the initial values of the global constants \code{x1} and \code{y1}
        and claims that no further choices need to be resolved to follow
        the execution to a violation of the postcondition.
        Scripts available at 
        the \href{https://gitlab.com/sosy-lab/benchmarking/sv-lib}{\Language} repo
        as \href{https://gitlab.com/sosy-lab/benchmarking/sv-lib/-/blob/format-1.0/examples/core-verification/loop-add-incorrect-ensures.svlib}{verification}
        and \href{https://gitlab.com/sosy-lab/benchmarking/sv-lib/-/blob/format-1.0/examples/core-validation/loop-add-incorrect-ensures.svlib}{validation} tasks}
    \vspace{-4mm}
\end{figure}

Finding bugs in programs is also an important
application of verifiers.
In the same manner in which we want to simplify the validation
of correctness proofs when compared to verification
from scratch, we also want to simplify
the validation of counterexamples compared
to verification from scratch.
Therefore, counterexamples fully describe a
single path through the program which leads 
to the violation of the property.
In particular, they resolve all types of non-determinism
relevant to uniquely characterize the path to the violation,
and provide the SMT model, such that the user
can execute the resulting trace through the program
and see the property violation. This ensures
that the validation can be done
through an interpreter-like execution of the program.

To show this in practice, 
let us consider~\cref{fig:verifying-incorrect-safety-property},
which is a slight
modification of the program in~\cref{fig:verifying-correct-program},
where we introduce an extra iteration of the
loop inside the function \terminal{add},
which results in a violation of the \terminal{:ensures}
annotation of \terminal{proc-add}.
An example witness showing a counterexample is
shown in~\cref{fig:incorrect-safety-witness}.
The witness presents a single path through the program,
by resolving all (implicit) non-determinism,
which in this case is only the non-deterministic
initialization of the global 
variables \terminal{x1} and \terminal{y1}.
Note that even though the local variable \terminal{y}
and output variable \terminal{x} are also non-deterministically
initialized by the procedure \terminal{add}, the
witness does not need to resolve this non-determinism,
since the values of these variables are fully determined
by the inputs \terminal{x1} and \terminal{y1} and the
assign statement at~\cref{line:incorrect-program-first-assign}. 
The trace ends with the full description of the violated tag,
encompassing its name and if applicable the violated term.

\subsection{Verifying an Incorrect Program for a Liveness Property}
\label{sec:verifying-incorrect-liveness-property}

\begin{figure}[t]
    \vspace{0mm}
    \hspace{2mm}
    \subfloat[Program with incorrect liveness property]{
      \parbox{0.52\textwidth}{
        \lstinputlisting[
                        language=SVLIB, 
                        numbers=left,
                        stepnumber=1,
                        firstnumber=1,
                        numberfirstline=true,
                        numbersep=2pt,
                        numberstyle=\tiny,
                        escapechar=@]
        {
            examples/simple-incorrect-liveness.svlib
        }
      }
      \label{fig:verifying-incorrect-liveness-property}
    }%
    \subfloat[Example witness for~\cref{fig:verifying-incorrect-liveness-property}]{
        \raisebox{-2.07cm}{
            \parbox{0.40\textwidth}{
                \lstinputlisting[
                            language=SVLIB, 
                            numbers=none,
                            stepnumber=2,
                            firstnumber=1,
                            numberfirstline=true,
                            escapechar=@]
                {
                    examples/simple-incorrect-liveness-witness.svlib
                }
            }
        }
        \label{fig:incorrect-liveness-witness}
    }
    \vspace{-1mm}
    \caption{\Language program violating a liveness
        property (left) and a possible witness for it (right),
        expressing a lasso that stays forever in the loop with \code{(= y 0)}.
        Scripts available at 
        \href{https://gitlab.com/sosy-lab/benchmarking/sv-lib}{\Language} repo
        as \href{https://gitlab.com/sosy-lab/benchmarking/sv-lib/-/blob/format-1.0/examples/core-verification/loop-add-incorrect-liveness.svlib}{verification}
        and \href{https://gitlab.com/sosy-lab/benchmarking/sv-lib/-/blob/format-1.0/examples/core-validation/loop-add-incorrect-liveness.svlib}{validation} tasks}
    \vspace{-4mm}
\end{figure}

In contrast to safety counterexamples,
which can be validated
by executing the program along the described path,
for the validation of liveness violations,
one requires reasoning about infinite traces.
This requires using abstractions like invariants
to describe the infinite part of the trace.
We express the infinite trace which violates the
liveness property by a finite prefix
followed by modular abstractions that guarantee
that the property of the annotated tag is violated.

As an example, consider the program
in~\cref{fig:verifying-incorrect-liveness-property},
which modifies the program
in~\cref{fig:verifying-correct-program}
such that \terminal{y} is decremented only
if it is larger than \terminal{1}.
This results in an infinite loop,
for any input where \terminal{y0} is 
larger than \terminal{1},
violating the liveness property
expressed by the \terminal{:not-recurring} tag.

The corresponding witness is shown
in~\cref{fig:incorrect-liveness-witness},
which describes a finite trace by
resolving all (implicit) non-determinism
which cannot be inferred from the program structure.
In this case, the only non-determinism
is the initialization of the global constants
\terminal{x1} and \terminal{y1}, since the
local variable \terminal{y} and output variable \terminal{x}
are fully determined by the inputs and the
program structure.
After the resolution of all non-determinism,
the witness ends with the violated tag.
Finally the infinite part of the trace
is described by a sequence of \terminal{using-annotation}
commands which express the inductive abstractions
required to proof that all paths from that state
fulfilling the annotations violate the tag.
For the validation of such a witness,
only the tag annotations 
inside of the \terminal{select-trace}
command can be used when traversing the infinite part
of the trace. In particular, this means that
if existing annotations are relevant, they also need to be
repeated inside of the \terminal{select-trace} command.

\subsection{Verifying a Program for an Insufficient Invariant}
\label{sec:verifying-incorrect-invariant}

\begin{figure}[t]
    \vspace{0mm}
    \hspace{2mm}
    \subfloat[Program with an insufficient invariant]{
        \parbox{0.52\textwidth}{
            \lstinputlisting[
                        language=SVLIB, 
                        numbers=left,
                        stepnumber=1,
                        firstnumber=1,
                        numbersep=2pt,
                        numberstyle=\tiny,
                        numberfirstline=true,
                        escapechar=@]
        {
            examples/simple-incorrect-invariant.svlib
        }
        }
        \label{fig:verifying-incorrect-invariant}
    }%
    \subfloat[Insufficient invariant witness]{
        \raisebox{-2.04cm}{
            \parbox{0.40\textwidth}{
                \lstinputlisting[
                                language=SVLIB, 
                                numbers=none,
                                stepnumber=2,
                                firstnumber=1,
                                numberfirstline=true,
                                escapechar=@]
                {
                    examples/simple-incorrect-invariant-witness.svlib
                }
            }
        }
        \label{fig:incorrect-invariant-witness}
    }
    \vspace{-1mm}
    \caption{\Language script with insufficient invariant (left) and a possible witness for it (right).
        The missing part is \code{(<= y 0)} so that \code{y}
        might still contribute to the sum of the invariant
        and thus the postcondition does not follow.
        Note that the \code{leap} step encodes a transition that was not part of the original program
        but rather reflects the overapproximation expressed by this imprecise invariant.
        Scripts available at 
        the~\href{https://gitlab.com/sosy-lab/benchmarking/sv-lib}{\Language} repo
        as \href{https://gitlab.com/sosy-lab/benchmarking/sv-lib/-/blob/format-1.0/examples/core-verification/loop-add-incorrect-invariant.svlib}{verification}
        and \href{https://gitlab.com/sosy-lab/benchmarking/sv-lib/-/blob/format-1.0/examples/core-validation/loop-add-incorrect-invariant.svlib}{validation} tasks}   
    \vspace{-4mm}
\end{figure}

It can occur that an invariant holds for all program executions,
but is not sufficient to prove the correctness of the program.
This might be the case for states that fulfill the invariant
but no path exists in an actual program execution that reaches such a state.
Then the invariant might be either not inductive or not safe.
Therefore, a counterexample for an invariant needs to
be able to enter an arbitrary state satisfying
the invariant and then provide a counterexample either
to the inductiveness of the invariant or to the specification.
The same holds also for contracts,
since they are required to be modular,
and for ranking functions,
since they are required to decrease in each iteration.

\Cref{fig:verifying-incorrect-invariant} shows a program
which is similar to the one in~\cref{fig:verifying-correct-program},
but annotated with an invariant which is not strong
enough to prove the post-condition \terminal{:ensures}
of procedure \terminal{add}.
This is because to proof the post-condition, one needs
to know that \terminal{y} is always
larger than or equal to \terminal{0}.
In the modular setting required for invariants,
where every variable modified by the loop is non-deterministically
chosen to fulfill the invariant at the loop-head,
we cannot prove the post-condition in this case.

\Cref{fig:incorrect-invariant-witness}
shows a possible witness for this.
It first resolves the (implicit) non-determinism
of the global constants and then calls the procedure to be verified.
Once it reaches the tag \terminal{while-loop},
it \terminal{leap}s into a state in which
the variables modified in the loop body are chosen arbitrarily
but satisfying the conjunction of all
\terminal{:invariant}s at that tag.
In this case a possible choice for the modified variables is
$x \mapsto 3$ and $y \mapsto -1$.
Note that it is possible to \terminal{leap}
to this state,
since it fulfills the conjunction of all \terminal{:invariant}s
even though it will never
occur in an actual execution of the program.
After \terminal{leap}ing, the witness provides
the violated tag, i.e., \terminal{:ensures}
for the tag \terminal{proc-add}.

\vspace{-1mm}
\subsection{Witness Validation}
\label{sec:witness-validation-example}

\begin{figure}[t]
    \vspace{3.5mm}
    \lstinputlisting[
                    language=SVLIB, 
                    numbers=left,
                    stepnumber=1,
                    firstnumber=1,
                    numberfirstline=true,
                    escapechar=@,
                    numbersep=2pt,
                    numberstyle=\tiny,
                    xleftmargin=4.3mm]
    {
        examples/simple-correct-validation-task.svlib
    }
    \vspace{-1mm}
    \caption{Validation task produced by combining
        the \Language~script from~\cref{fig:verifying-correct-program}
        with the witness from~\cref{fig:correct-program-witness};
        script available at 
        \href{https://gitlab.com/sosy-lab/benchmarking/sv-lib}{\Language} repo
        as \href{https://gitlab.com/sosy-lab/benchmarking/sv-lib/-/blob/format-1.0/examples/core-validation/loop-add.svlib}{validation task}}
    \label{fig:correct-program-validation-task}
    \vspace{-5mm}
\end{figure}

Once we have obtained a witness for a verification task,
we can construct a validation task from the verification task and the witness
by inserting all commands of the witness (i.e., the complete witness
without the wrapping parentheses)
just before the \terminal{verify-call} which produced it.
For example, for the correct program
in~\cref{fig:verifying-correct-program} and
the corresponding witness in~\cref{fig:correct-program-witness},
this results in the \Language script in~\cref{fig:correct-program-validation-task},
which represents a new verification task.
This is one of the major advantages of having both
verification tasks and witnesses
in the same format of \Language~scripts,
since witness validation can be trivially reduced
to verification.
In general, the concatenation of two \Language~scripts
is also a \Language~script.

\section{\Language{} Commands}
\label{sec:commands}

An SMT-LIB script consists of a sequence of top-level commands
that are processed sequentially by the solver.
Commands encompass declaration, definitions, and axioms
to describe the problem,
as well as queries that probe for solutions to the problem,
such as satisfiability, models, and proofs.
This principle allows for incremental sessions
in which a series of queries is posed by the client and answered
by the solver in a dialogue.
This execution mode is a significant difference to
model checking, which processes a whole
verification task at once, not
having an active connection where commands can
be added after a response.

The design of \Language~follows the same principle:
It extends SMT-LIB by top-level commands to declare global variables,
to define (recursive) procedures and desired correctness properties.
In particular, it therefore inherits sorts, identifiers/symbols, 
and terms from SMT-LIB.
The verification tool can then be queried for whether
a specific procedure call is correct.
If a verdict has been reached, the verifier can be instructed
to return a witness (either for correctness or for violation).
Being an extension, \Language~therefore inherits many
elements from SMT-LIB, for example some commands, terms, and sorts.

\Cref{fig:commands} shows the syntax of all 
top-level commands in \Language.
The list of SMT-LIB commands which
need to be supported is given in~\cref{ssec:smt-lib-commands}.
Apart from these commands, all lines starting with a semicolon
are treated as comments and ignored by the verifier.

\begin{figure*}[t]
    \centering
    \begin{minipage}{0pt}
        \input{figures/commands}
    \end{minipage}
    \caption{Syntax of top-level commands in \Language}
    \label{fig:commands}
\end{figure*}

Whenever the verifier finished processing a
command, or an external interruption of the verifier was 
triggered, like receiving a ``SIGTERM''
it should provide a response.
In particular, the verifier should always formulate a
response to the last command before terminating,
in case it is interrupted externally.
For each command, except explicitly 
stated otherwise like for \terminal{verify-call},
the appropriate response is
\terminal{success} if the option \terminal{:print-success}
is enabled as in SMT-LIB or an empty response if it is not enabled.
In case of an error,
for example if the program is not well-formed
or some features are not supported,
the verifier should respond with \terminal{error}.

\subsection{Global Variables}

The command $\LP \terminal{declare-var}\ x\ \sigma \RP$
introduces a global program variable~$x$ of sort~$\sigma$.
It is analogous to SMT-LIB's \terminal{declare-const}
but global program variables may occur on the 
left-hand side of assignments
and thus may change their respective value throughout computations.
Global variables implicitly come in 
scope of all subsequent procedure declarations ---
the use of global variables inside a procedure body does not need to be declared explicitly.

The declaration of a global variable immediately
initializes the variable to an arbitrary value
of the corresponding sort.
Therefore, reading a declared variable which has
not yet been assigned any value is defined behavior.

One important purpose of global variables is to model
state parts that are implicit in high-level languages
but explicit in logical encodings, such as the state of the heap.
Additionally, global variables are a concept that is heavily relied on
by typical verification approaches 
and therefore we support them first-class.

\subsection{Procedure Declarations}
\label{ssec:define-proc}

The command
$\LP \terminal{define-proc}\ \rho\
    \LP \mathit{in} \RP\
    \LP \mathit{out} \RP\
    \LP \mathit{local} \RP\
    s
    \RP$
introduces a procedure with name~$\rho$,
a list of formal input parameters~$\mathit{in}$,
a list of formal output parameters~$\mathit{out}$
(return values),
and a list of local variables~$\mathit{local}$.
The parameters and variables are available in the body statement~$s$ of the procedure.
The syntax of statements is explained in \cref{sec:statements}.
Each output parameter and local variable has an 
arbitrary initial value of its corresponding sort
whenever the procedure is called,
but they can be assigned to in the body of the procedure.
There is no explicit return of output values; the last
value assigned to an output parameter
is considered the return value of the procedure
for that output parameter.

In their body statements, procedures are allowed to use only variables and
procedure calls that
have been defined before.
Additionally, the procedure body cannot modify
the values of its input parameters.

\subsection{Recursive Procedures}

Since commands are processed sequentially,
normally all symbols used in a procedure
declaration need to be known at the time
of its declaration.
This does not work for recursive procedures,
since they need to reference the procedure 
symbol introduced
by the procedure declaration before the command
has been processed.

The command $\terminal{define-procs-rec}$
defines a set of recursive procedures.
It takes a non-empty list of procedure
declarations (see~\cref{ssec:define-proc})
which can reference each other.
It first defines the signature of each procedure
(i.e., name, input parameters, output parameters, local variables)
and only once this is known, it introduces the bodies
of the procedures, which can reference each other.
This makes the parsing easier, since all signatures
are known before parsing the bodies.

For now, we do not actually require the procedures to be recursive,
therefore it is fine to just embed all procedures in 
the task into a single such command.
However, both verifiers as well as benchmark generators
are encouraged to either recognize or limit such cases.

\subsection{Annotating Tags}
\label{ssec:annotate-tag}

The command
$\LP \terminal{annotate-tag}\ \tau\ a_1, \dots, a_n \RP$
links program locations~$\tau$ to their specification,
which are expressed by attributes~$a_1, \dots, a_n$.
Here, $\tau$ is a \emph{tag} potentially mentioned as part of the
statements in the procedure bodies, identifying locations of interest.
If the tag $\tau$ does not exist, the annotation command is ignored.
The syntax of these attributes follows the generic
annotation mechanism of SMT-LIB with keywords $\terminal{:kw}$
or keyword-value pairs $\terminal{:kw}~e$ for some S-expression~$e$.
The properties supported by 
\Language~are described in \cref{sec:properties}.

Program variables appearing in an attribute
are interpreted in the context of the beginning of the statement 
that is tagged (not in the context of the \terminal{annotate-tag} command).
This means that the sorts of the
variables, and the variable referenced (input/output/local of the procedure or global) 
by the variable identifiers in the property must be resolved when 
interpreting the \terminal{annotate-tag} command.
See also the discussion on well-formedness of programs in \cref{sec:well-formed-programs}.

SMT-LIB sorts, constants, and function symbols appearing in an attribute are interpreted
at the location of the \terminal{annotate-tag} command.
This implies, for example, that witnesses can introduce new definitions that were not available
in the original verification task.
It also means that some care has to be taken when moving such annotations
between the global command level and inline attributes of statements.
This condition can be detected syntactically
and if needed the corresponding function definitions can be moved to an earlier location in the script.

\subsection{Selecting Traces}
The command
$\LP \terminal{select-trace}\ \pi \RP$
selects a concrete trace~$\pi$
through the program for the (prefix of) possible executions.
The trace resolves initialization of global variables,
specifies the entry-point into the execution,
and resolves all non-deterministic choices 
including initialization of local variables,
as described in \cref{sec:witnesses}.
Furthermore, it can represent jumps into
arbitrary states, and set tags to
represent infinite loops.
The full description of traces is given
in~\cref{ssec:counterexample-safety}
for safety violations, in~\cref{ssec:counterexample-side-conditions}
for violations of modular specifications,
and in~\cref{ssec:counterexample-liveness} for 
violations of liveness properties.

The primary goal of this command is to
restrict the verification to a single
trace for witness validation.
Therefore, only $\terminal{verify-call}$ commands come after the 
command $\terminal{select-trace}$; any other command is ignored.

\subsection{Verification Queries}
The command
$\LP \terminal{verify-call}\ \rho\ \LP t_1, \dots, t_n \RP \RP$
is the analogue of SMT-LIB's $\LP \terminal{check-sat} \RP$.
It instructs the verifier to check that all executions starting
at the specified entry procedure~$\rho$
satisfy the properties annotated as part of the program, i.e.,
specified by prior \terminal{annotate-tag} commands.
The initial states are given by the interpretation of the
terms~$t_1, \dots, t_n$.
They denote the (possibly symbolic) values of input parameters of~$\rho$,
possibly using global variables,
all of which are described by
prior definitions and constraints
from~$\LP \terminal{assert}\ \phi \RP$ commands.
In case a \terminal{select-trace} command
is present as the last command before
the \terminal{verify-call} command,
the verification is restricted to the
single trace specified by the
\terminal{select-trace} command.

The \terminal{verify-call} command poses the question:
\emph{Is the \Language script correct with respect to its specifications,
when starting all executions at the entry procedure~$\rho$ with the given arguments?''}
The response to a \terminal{verify-call} command
is one of \terminal{correct}, \terminal{incorrect}, \terminal{unknown}, \terminal{unsupported}, and \terminal{error}.
The verifier produces \terminal{correct}
if it was able to verify that the program
satisfies \emph{all} specified properties,
\terminal{incorrect} if it
found a counterexample to \emph{at least one} 
of the specified properties,
\terminal{unknown} if it was unable to reach a conclusion,
\terminal{unsupported} if it does not support some feature of the \Language~script, and
$\LP \terminal{error} \dots \RP$ if an error was encountered
during the verification process (same format as in SMT-LIB).

If the verification task contains one or multiple \terminal{select-trace} commands for that procedure,
the verifier may make use of them to see if they do in fact violate the respective property.
This involves checking which of these apply to a specific call site,
wether the trace exists in the program,
and whether the indicated property violation can be confirmed.
This is of interest particularly for the construction of tasks for validation of violation witnesses.
In that case, the verdict should still be \terminal{incorrect}.
If some of those traces are erroneous, this can be reflected in the response as well.
More details on this are in \cref{sec:witnesses}.

\subsection{Witness Retrieval}

Whenever a verification query is answered with \terminal{correct}
or \terminal{incorrect},
the client may request a witness as evidence
in form of a correctness or a violation witness,
with the command $\LP \terminal{get-witness} \RP$.
The format of the witnesses is described in \cref{sec:witnesses}:
it consists of structured data.
Since the response may span multiple lines, it is enclosed by a pair of parentheses.

The command $\LP \terminal{get-witness} \RP$
requests the verifier to return a witness
for the last verification query.
If any other command was executed between the
last verification query and the \terminal{get-witness} command,
the command should produce an error, since the
previous witness may be invalid, for example, if
the inserted command is an \terminal{annotate-tag} command.

In case witness production is not enabled,
the \terminal{get-witness} command should respond with an error.
Additionally the output channel to write the witness to
defaults to \texttt{stdout} if not specified otherwise.

\subsection{Tool-Specific Calls}
Tools may introduce further specific queries,
for example to run a concrete or symbolic simulation,
to generate best-effort invariants 
independently of a specification,
or to generate test suites.
We suggest such extensions to follow the same scheme
as the function \terminal{verify-call}.
Namely, they should specify the procedure to call
and the arguments to pass.
If more is necessary, it should be encoded using
the already known commands.

While not strictly required, we recommend that
responses to tool-specific calls adhere to the
design principles of \Language.
In particular, we recommend that the response
be either a single symbol like \terminal{correct} and \terminal{incorrect} 
or a sequence 
of \Language~commands like the result of the 
\terminal{get-witness} command.
This ensures easier reuse of the commands,
and may allow them to be added to the
\Language~standard in the future.

\subsection{SMT-LIB Commands}
\label{ssec:smt-lib-commands}

\begin{figure*}[t]
    \centering
    \begin{minipage}{0pt}
        \input{figures/smt-lib-commands}
    \end{minipage}
    \caption{
        SMT-LIB commands allowed in \Language~scripts
        (taken from~\cite{SMTLIB27})
    }
    \label{fig:smt-lib-commands}
\end{figure*}

While many SMT-LIB commands make sense in the
context of \Language, others do not, 
like \terminal{check-sat}.
\Cref{fig:smt-lib-commands} lists all SMT-LIB commands
supported by \Language.
Their usage is defined over syntactic elements, which are
part of SMT-LIB; please refer to the SMT-LIB
standard~\cite{SMTLIB27} for their syntax and semantics.

The purpose of including definitions is to model the application domain
or the semantics of a higher-level language from which the verification task has been compiled,
for example to model heap data structures or similar.
We have deliberately included the \terminal{assert} command as well,
since not all features of such background theories are easily described by proper definitions,
rather one may want to resort to axiomatization of uninterpreted functions instead.
One example is the axiom $a \not\in s \implies |s \cup \{a\}| = |s| + 1$, for a finite set~$s$.
Because sets are not freely generated (they admit non-unique syntactic presentations of semantically equivalent values),
there is no obvious definitional principle that is recursive over the elements.
Another example is the axiomatic under-specification of trigonometric functions,
for which precise reasoning is difficult for an SMT solver.
In general, front-end tools like Dafny or Frama-C often ship with large preludes encoded in this way,
which we want to accommodate as well.

Note that axioms can render the verification meaningless: for example, adding 
$\LP\terminal{assert}\ \terminal{false}\RP$ to the script should make the program vacuously correct.
We consider this as an orthogonal problem that is not in scope for \Language verifiers.

However, there is an important restriction:
Formulas~$\phi$ asserted by $\LP\terminal{assert}\ \phi\RP$ can only constrain
the SMT model but \emph{not} the dynamic execution state of the program.
Thus, $\phi$ must not contain references to  global variables declared with \terminal{declare-var}.
This implies that constraints on the initial global state at \terminal{verify-call}
must be specified by other means, such as \terminal{assume} statements inside the procedures.

\section{\Language{} Statements}
\label{sec:statements}

\begin{figure*}[t]
    \centering
    \begin{minipage}{0.9\textwidth}
        \input{figures/statements}
    \end{minipage}
    \caption{
        Syntax of statements. 
        Each colored letter represents a distinct fragment of \Language;
        \basicLanguageFragment{} corresponds to basic statements,
        \procedureStatementsFragment{} to procedure calls and returns,
        \unstructuredControlFlowFragment{} to unstructured control flow,
        \structuredControlFlowFragment{} to structured commands,
        and \nondeterministicFragment{} to nondeterminism;
        the superscript $^*$ indicates an extension of the respective fragment
        only under certain conditions
    }
    \label{fig:statements}
\end{figure*}

Statements occur as part of procedures (see~\cref{sec:commands})
to define computational steps.
In the design of \Language,
we aim to standardize constructs that 
are well-understood and commonly used,
without imposing a particular encoding to the program.
For instance, the language has unstructured 
control flow but also loops,
acknowledging that verification tools 
and approaches may prefer one over the other.
We express this by using \emph{fragments}, 
akin to SMT-LIB logics, for different feature sets.
In some cases it is possible to 
translate back and forth between fragments,
for example by replacing loops via gotos,
or by inlining non-recursive procedures.
Therefore, in the absence of support for certain features,
verifiers may choose to translate to a supported fragment.

\Cref{fig:statements} shows the syntax of statements
with the respective fragments being indicated on the right.
The intended semantics of programs are 
finite and infinite sequences of states,
which in turn are comprised of global and local variables.
For safety proofs, binary relations between states
that collapse the (in)finite sequences are an alternative view.
We do not define the semantics here but we will present 
verification conditions in a future version of this document,
which explain the logical properties of statements precisely.

Many statements make use of symbols, sorts and/or terms
in their definition.
These are SMT-LIB symbols, sorts and terms, we therefore
delegate the description of their syntax and
semantics to the SMT-LIB standard~\cite{SMTLIB27}.
In practice whoever SMT-based verifiers will not
need to understand them, since they can be delegated
to the underlying SMT solver in most cases.
One exception to this is that symbols starting
with \reservedSymbolPrefix{} cannot appear in
\Language~scripts, even though they are allowed
in SMT-LIB. This allows verifiers to easily be able
to create new variables by pre-fixing~\reservedSymbolPrefix{}
to an existing variable.
These symbols, are also not allowed to appear in 
the witness, so verifiers need to take care to
rename such symbols when generating witnesses.

\subsection{Basic Language (\basicLanguageFragment)}
The basic language fragment (\basicLanguageFragment) encompasses 
assumptions, assignments,
as well as sequential composition
and annotations.

\inlineheadingbf{Assumptions}
Assumptions
$\LP \terminal{assume}~\phi \RP$
restrict the possible executions to those
satisfying the boolean term~$\phi$.
By convention, $\LP \terminal{assume}\ \code{true} \RP$
is used to encode the statement that takes a step but does nothing,
$\terminal{skip}$.
Analogously, $\LP \terminal{assume}\ \code{false} \RP$
denotes that the current program execution 
should no longer be considered at all.
In particular, this means that 
$\LP \terminal{assume}\ \code{false} \RP$
always terminates, but has no final state~\cite{DataRefinementMiracles}.

Of course, \terminal{assume} is a intended modeling construct that
cannot be executed in general.
Moreover, even though the intention of such assumptions is to constrain
the program's execution, it may have some effect on the interaction of the underlying SMT model
with possible states of the program.
This implies that the models that initialize the state as part of counterexample traces
are implicitly required to reflect such assumptions.
This is intended, for example, when the execution depends on an uninterpreted global constant,
such as the length of some array data structure or the assumption of bounded resources.

\inlineheadingbf{Assignments}
Parallel assignments
$\LP \terminal{assign}\
    \LP x_1\ t_1 \RP\ \dots\
        \LP x_m\ t_m \RP$.
evaluate the right-hand side terms~$t_1, \dots, t_m$ first
in that order
and then simultaneously and atomically replace
the values of~$x_1, \dots, x_m$ with the results.
An assignment is well-formed if the variables are distinct,
and the right-hand side types match the corresponding variable types.

Note that neither constants nor input parameters of procedures
can be assigned to, and therefore cannot appear on the left-hand 
side of assignments.

Furthermore, note that SMT-LIB functions can be assigned to variables,
but procedures cannot, since we want all their calls to be explicit,
as a statement and not occur inside of terms, as in traditional programming
languages.
Since allowing procedure calls in terms introduces 
a lot more complexity for verification tools,
since they need to somehow decide in what order to handle
the calls, and the evaluation of terms would no longer be side-effect free.

\inlineheadingbf{Sequential Composition}
Sequential composition
$\LP \terminal{sequence}\ s_1\ \dots\ s_n \RP$
executes the respective constituents in sequence in the
given order.
Another way of writing \terminal{skip} is to
take an empty list of statements in
a \terminal{sequence}.
This is desired, in order to simplify the
translation of programs into \Language.

\inlineheadingbf{Annotations}
The basic language furthermore supports the annotation mechanism
from SMT-LIB at the level of statements:
$\LP \terminal{!}\ s\ \dots \RP$
denotes that the command~$s$ has additional meta-data attached.
There are two-predefined annotations types, tags and properties.
For an explanation about allowed properties see\cref{sec:properties}
Note that annotations are compositional, i.e.,
$\LP \terminal{!}\ \LP \terminal{!}\ s\ a \RP\ b \RP$
is the same as $\LP \terminal{!}\ s\ a\ b \RP$.

Tags $\terminal{:tag}\ \nonterminal{symbol}$,
allow users and front-end tools to mark the (entire) statement~$s$
as being of interest to a concern related to the given symbol.
It is allowed that the same tag can be attached to 
multiple places in the program,
for example to mark groups of locations that 
should satisfy a joint property.
The primary use case of tags is to link program locations to properties
via the \terminal{annotate-tag} command.
We emphasize that the \terminal{:named} annotation
from SMT-LIB or other conventions can be used
to (uniquely) identify program locations for other purposes,
such as the encoding of structured loops 
into explicit control flow (see below).

Properties
are used to express properties of the program.
For example,
$\LP \terminal{!}\ s\ \terminal{:check-true}\ \phi \RP$
specifies that whenever~$s$ is about to execute,
formula~$\phi$ should hold, i.e., $\phi$ is a location invariant
at the start of~$s$.
With the use of tags, we can liberally and losslessly
move such property annotations in and out of programs.
Note that when $\phi$ is moved out of the program by
use of tags, and a \terminal{annotate-tag} command,
all symbols are resolved in the context at the 
beginning of the 
tagged statement~$s$.

The distinction between what information is part of statements
and what is part of the annotation/property language
is driven by the principle that
annotations should never be used to describe the semantics of the statements.
In particular, tags are orthogonal to labels, 
where the sole purpose of the latter
is to identify jump targets.
This also explains, why assumptions are realized
by \terminal{assume} statements,
as they are an essential construct to constrain executions
in relation to non-deterministic assignments and choice,
whereas assertions are realized separately as
\terminal{:check-true} annotations,
as these reflect properties to be proven
in relation to the executions.

In order to export witnesses correctly, it is important
that certain crucial statements are marked with script unique tags.
These are loops, labels and the outermost statement in a procedure body.
Ensuring this is the responsibility of the script generator,
not of the solver.

\subsection{Procedures (\procedureStatementsFragment)}
The procedural fragment (\procedureStatementsFragment) of \Language 
comprises statements
for procedure calls and for returning from them.

\inlineheadingbf{Procedure Call}
A procedure call can be made using
$\LP \terminal{call}\ \rho\
    \LP t_1\ \dots\ t_n \RP\
    \LP y_1\ \dots\ y_m \RP\
    \RP$.
This statement calls the procedure $\rho$, which has to have been introduced
by the \terminal{define-proc} command previously.
The terms $t_1, \dots, t_n$ denote the values of the input parameters,
and the variables~$y_1, \dots, y_m$ receive the output values.
A call is well-defined if the respective number
and types of input/output arguments match the procedure definition.
Moreover, similarly to the assignment statement,
the variables~$y_1, \dots, y_m$ need to be distinct.

The semantics of calls is stack-oriented as expected, i.e.,
each invocation of a (recursive) procedure gets its own copies of
the input/output/local variables declared by the procedure.
Furthermore, only values are passed from the input arguments to the procedure,
and from the procedure back to the output arguments.
This means that even if a constant is returned from the function,
the caller can assign its value to a variable which can be modified.
This means that a symbol being constant or not does not affect its
assignability with respect to procedure calls.

A procedure returns either when a \terminal{return} statement is executed,
or when the end of the procedure body is reached.
When returning, the current values of the output variables
are passed back to the caller.

\inlineheadingbf{Return}
The second part of the procedural fragment is
the \terminal{return} statement.
Note that this statement itself does not
denote the values of the output parameters of the surrounding procedure,
rather, these are given by ordinary assignments,
so that the outputs produced are whatever values
are currently stored inside the output variables.

\subsection{Unstructured Control Flow (\unstructuredControlFlowFragment)}

The fragment of unstructured control 
flow (\unstructuredControlFlowFragment)
is comprised of \terminal{label}s and \terminal{goto}s
with the standard meaning.
In case the unstructured control flow
fragment is used without the structured
control flow fragment,
a restricted form
of conditional jumps 
is automatically included into the unstructured control flow fragment.

\inlineheadingbf{Labels}
Labels $\LP \terminal{label}\ \ell \RP$
mark positions in the program that can be jumped to
by \terminal{goto} statements.
Labels~$\ell$ 
must be unique within
each procedure, such that the target of a jump
is always unambiguous.
Labels are separate constructs from tags,
therefore it is possible to have tags and labels
with the same name in the same procedure.
This may be even desirable to link properties to
the label locations.

\inlineheadingbf{Goto}
Goto statements
$\LP \terminal{goto}\ \ell \RP$
jump to the position marked by label~$\ell$.
Jumps can only be made within the same procedure,
this guarantees that all variables which were in scope
at the call-site are still in scope at the target location.
In case a label is not found in the current procedure
matching the target of a \terminal{goto},
the program is considered ill-formed.

\inlineheadingbf{Conditional Jumps}
In case the unstructured control flow fragment
is used without the structured control flow fragment,
conditional jumps
$\LP \terminal{if}\ \phi\ \LP \terminal{goto}\ \ell \RP \RP$
are supported as well.
These statements behave the same way as
an \terminal{if} statement.
They are required whenever the structured control flow fragment
is not used, to allow for branching behavior.

\subsection{Structured Commands (\structuredControlFlowFragment)}
The fragment of structured control flow (\structuredControlFlowFragment)
includes the typical programming 
constructs \terminal{if}, and \terminal{while}. 
In order to support a well-defined translation of
\terminal{if} and \terminal{while}
statements into unstructured control flow,
a useful convention is to annotate these
with \terminal{:named} attributes,
which gives them a (unique) name in addition 
to tags and labels present,
from which the labels of entry and 
exit locations can be generated.

\inlineheadingbf{Conditions}
For \terminal{if} statements, the absence of 
the second statement in a conditional defaults to
$\LP \terminal{assume}\ \terminal{true} \RP$ as usual.

\inlineheadingbf{Loops}
Only \terminal{while} loops are supported, since
they are sufficient for all cases requiring other types of loops.
Tags resp. properties attached as annotations to loops specifically
can identify high-level proof principles such as loop invariants
for safety properties
and well-founded orders for termination properties.

\inlineheadingbf{Break and Continue}
Statements \terminal{break} and \terminal{continue}
are only allowed to occur inside the scope of a loop.
They have the standard meaning of
exiting the innermost surrounding loop immediately,
and skipping to the next iteration of the innermost 
surrounding loop, respectively.

\subsection{Nondeterminism (\nondeterministicFragment)}
For modeling purposes, nondeterminism (\nondeterministicFragment) is
a key feature of intermediate languages.
We consider two types of nondeterminism,
data nondeterminism through \terminal{havoc},
and control-flow nondeterminism through \terminal{choice}.
Note that control-flow nondeterminism can be encoded
via data nondeterminism whenever the structured control-flow fragment
is available, by using a nondeterministic variable
in combination with \terminal{if} statements.

All nondeterminism in \Language~should be interpreted
as potential program behaviors which might occur during execution
and hence should be considered during verification.
In that regard, all choice is ``demonic''.
Albeit the distinction between demonic and angelic choice
only makes sense with respect to a game-like semantics,
which definitely goes beyond the scope of \Language.
In concrete counterexample traces, choices 
for \terminal{havoc} are resolved to values,
similarly, the branch taken by \terminal{choices} 
must be made explicit.

\inlineheadingbf{Havoc}
The atomic statement
$\LP \terminal{havoc}\ x_1\ \dots\ x_n \RP$
overwrites the values of (distinct)
variables $x_1, \dots, x_n$ by fresh values.
This is done atomically and simultaneously.

\inlineheadingbf{Choice}
The statement
$\LP \terminal{choice}\ s_1\ \dots\ s_n \RP$
selects one sub-statements among the $s_1,\dots,s_n$
to execute.
Note that Dijkstra's guarded commands can be written
by a combination of $\terminal{choice}$
and $\terminal{assume}$ statements.

\section{\Language{} Properties}
\label{sec:properties}

As indicated earlier, properties are kept 
distinct from the definition of the execution steps.
The intention is to clearly separate the two 
concerns of (a) modeling a system
and (b) specifying its desired behaviors.
Specifically, for \Language, we aim to avoid 
encoding specification concerns
into the program definition, neither in 
the form of annotations like
assertions, contracts, or invariants~\cite{PenAndPaperToIndustrialTools},
nor in the form of, e.g., automata-theoretic 
constructions like monitoring automata
that are composed with the program~\cite{BLAST}.
The view is that such constructions reflect 
details of a specific verification approach
and therefore such artifacts
should be regarded as \emph{output}
of a proof procedure instead of being its input.

The design of linking specifications into programs 
is guided by this view.
As shown earlier, specifications are attached 
as properties to statements via 
the annotation mechanism borrowed from SMT-LIB.
This connection can either be drawn 
explicitly or alternatively,
indirectly by \terminal{:tag} annotations,
which are used to identify locations 
in the program that are linked
to properties by \terminal{annotate-tag} top-level commands
(or further mechanisms).

\begin{figure*}[t]
    \centering
    \begin{minipage}{0.9\textwidth}
        \input{figures/properties}
    \end{minipage}
    \caption{
        Syntax of property declarations;
        the fragment which introduces the need
        for this property, together with their
        classification as liveness \livenessProperty{}
        or safety \safetyProperty{} properties is indicated
        on the right-hand side
    }
    \label{fig:properties}
\end{figure*}

In its current format, the properties which
can be specified about~\Language~scripts are
only LTL properties.
Therefore, they are partitioned into
safety~(\safetyProperty) and
liveness~(\livenessProperty) properties.
Safety properties encode that something
bad never happens,
while liveness properties ensure that
something good eventually happens.
In this section, we describe the properties,
shown in~\cref{fig:properties}.
In general, we make use of relational terms,
which extend SMT-LIB terms by allowing the use of
certain constructs at each location a 
variable symbol is allowed.

\inlineheadingbf{Relational Terms}
Relational terms are constructed in the same
manner as terms, with the addition
of $\LP\terminal{at}~\nonterminal{symbol}~\nonterminal{symbol}\RP$,
which can be used in every location a symbol is allowed. 
It references the
value of a symbol at the beginning 
of the last statement visited which is marked
with the tag referenced by the
\terminal{at}.
The first parameter corresponds to
the variable whose value is referenced,
and the second parameter
corresponds to the tag.
This feature can be used to encode for example
the specification construct \terminal{old} in
languages like ACSL~\cite{ACSL}, 
JML~\cite{JML00}, or Dafny~\cite{Dafny} by
referencing the tag at the beginning of
the procedure body. This tag must always
exist in well-formed programs,
see~\cref{ssec:fully-well-formed-programs}.

The \terminal{at} construct may be applied to
symbols representing program variables only,
but not to general terms.
This avoids assigning meaning to nested \terminal{at} constructs
and avoids the need to keep track of bound variables in
terms, for example
$\LP\terminal{forall}~\LP\LP\terminal{k}~\terminal{Int}\RP\RP~\LP\terminal{at}~\LP\terminal{select}~\terminal{a}~\terminal{k}\RP~\tau\RP\RP$,
where bound variable~$\terminal{k}$ in the scope of \terminal{at} should \emph{not}
be resolved with respect to the state at the earlier occurrence of tag~$\tau$.
Front-end tools for \Language are encouraged to support
more general compositions using \terminal{at}
by compiling them into this lower-level form.

\subsection{Location Invariants (\safetyProperty)}

The basic safety property consists of a location invariant,
denoted by the attribute \terminal{:check-true}.
While in most programming languages location invariants are
characterized by assertions, we choose to use \terminal{:check-true}
instead in order to make the distinction between the SMT-LIB
\terminal{assert} and location invariants clearer.
A location invariant is always evaluated in the state before executing
the statement it is attached to, with the exception of \terminal{while}
loops, and \terminal{label} statements.
Counterexamples to location invariant violations are
expressed by counterexamples to safety, see~\cref{ssec:counterexample-safety}.

Some clarification should be made about the interpretation
when~$s$ is a $\LP \terminal{while} \dots \RP$ statement.
Should~$\phi$ be checked only once or on every iteration?
Arguments can be made in favor of both views,
but we find it more pragmatic to repeat the check each time when the
loop condition is evaluated.
The reasoning why this is the better option is as follows:
First, it gives a convenient way to specify semantic loop invariants
(without requiring them to be inductive).
Second, when translating loops into goto programs or CFA representation,
the annotation can simply be attached to the loop head
without the need to distinguish the first visit from later ones.
The same arguments apply for \terminal{label} statements,
where we decide that the assertion should be checked 
each time the label is jumped to.

\subsection{Invariants (\safetyProperty)}
\label{ssec:invariants}

For (un)structured control-flow, we support invariants
using the attribute \terminal{:invariant}.
Invariants are required to be fulfilled, the first time
the loop is entered, inductive, and be strong
enough to prove the desired properties if the
loop is abstracted by the invariant.
This applies to both structured \terminal{while} loops
and unstructured loops using labels and gotos.

Similarly to loops, annotating loop-heads, i.e., labels 
with \terminal{:invariant}s,
can help break cycles in such unstructured programs
to achieve modular verification.
There are multiple interpretations for where the invariants
should be in order to break cycles,
at least they should be at one node inside each cycle,
and at most at each label.
Due to these interpretations not providing clear
advantages over each other, we leave the
precise choice to the tools generating
or processing \Language~scripts,
and aim to provide a more formal definition
in a future version of the format.
In particular, if a verifier does not produce
invariants for labels not considered 
loop-heads/contributing to cycles,
invariants can be generated for them
from other existing annotations using
strongest post-conditions and weakest pre-conditions.

We use a weak notion of inductivity, which is with respect
to all the variables being modified within the loop.
This means, that if we start in the state where the loop
is entered the first time, then \terminal{havoc}
all modified variables inside the loop, execute
the body of the loop, then the loop invariant must hold.
In particular, this means that variables which are not modified
inside the loop, will retain the same value as
before the loops execution even when abstracting the loop using the
invariant.
This includes, but is not limited to, global variables,
and constants.

Note that finding out which variables are modified
inside the loop is purely a syntactic task, since
simply all variables on the left side of
an assignment,
or as output of a procedure call inside the loop body,
are considered to be modified.
This is possible since \Language~does not have a heap,
for which alias analysis would be required.
In particular, for arrays, we consider the whole array
to be modified, regardless the precise
locations written to.
In this case, the invariant must express
through the use of quantifiers
and relational terms that certain elements
of the array retain their value.

Whenever the over-approximation given by the invariant
is applied, for example, during inductiveness constraint,
\terminal{continue} and \terminal{break} statements
need to be handled correctly.
When a \terminal{continue} statement is executed,
then the invariant must hold again, producing no
further verification conditions for the continued
execution after the loop.
In contrast, when a \terminal{break} statement
is executed, then the invariant does not need
to hold anymore, and the execution continues
after the loop body.
In particular, this means that abstracting a loop
using its invariants can generate multiple
successor states, depending on the number of
\terminal{break} statements inside the loop body.

Having three properties which need to be proven
correct, for an invariant to be valid, there are
also three types of counterexamples which are possible.
If the invariant does not hold when the loop is entered
the first time, a counterexample to safety, see~\cref{ssec:counterexample-safety}
must be exported, like for assertions.
If the invariant is not inductive, a counterexample
to modular specifications must be
exported, see~\cref{ssec:counterexample-side-conditions}.
Finally, if the invariant is not strong enough
to proof the rest of the program correct,
the same type of counterexample as for inductivity must be exported.

\subsection{Statement Contracts (\safetyProperty)}

Procedure contracts are represented in terms of more general
statement contracts, defined by the pair 
of attributes \terminal{:requires}
and \terminal{:ensures}.
The \terminal{:requires} attribute
denotes the pre-condition of the statement,
while the \terminal{:ensures} attribute
denotes the post-condition of the statement.
The \terminal{:requires} attribute is evaluated
in the state directly \emph{before} executing
the statement it is attached to,
while the \terminal{:ensures} attribute
is evaluated in the state directly \emph{after} executing
the statement.

Statement contracts are checked inductively similarly
to invariants:

First, the pre-condition given by the \terminal{:requires}
attribute must hold in the initial state before the statement is executed.
Such preconditions can be violated in the same way as \terminal{:invariant}s
by a trace potentially containing \terminal{leap} steps
to abstract an arbitrary number of recursive calls towards a the violation~\cref{ssec:counterexample-side-conditions} when the contract
is attached to the top-level statement of a procedure.%
    \footnote{
We leave the use of these annotations
for loop contracts~\cite{LoopVerificationContractsInvariants}
for future work, since these are not widely-used.
The rationale is that within loop contracts,
referencing variables should be consistent across pre- and postcondition,
i.e., it would be preferential to let \code{x} in \terminal{:ensures}
annotations to refer to the arbitrary intermediate execution, not the final one state,
which is incompatible with the established convention for procedure contracts.
Unfortunately, $\LP \terminal{at}\ \code{x}\ \ell\RP$ does not have the right semantics to encode the reference the intermediate state
when~$\ell$ is the tag annotated to the loop.}

Second, the contract must have a modular proof with respect to
the surrounding context and the statically-determined set of
local/global variables modified within the statement.
This means that whenever the statement is executed with all variables,
which the statement modifies, being nondeterministic
and the chosen variables fulfill the \terminal{:requires} attribute,
then at the end of the statements execution, the \terminal{:ensures} 
attribute must hold.
Moreover, in the context of recursive procedures,
the statement contract can make use of inductive hypotheses about recursive calls as usual.

Third the post-condition given by the \terminal{:ensures} 
attribute must be strong enough to prove the remainder
of the program correct.
Whenever this does not hold a counterexample 
to modular specifications must be exported,
see~\cref{ssec:counterexample-side-conditions}.

\subsection{Ranking Functions (\livenessProperty)}

Ranking functions, expressed by the attribute \terminal{:decreases},
can be used to prove termination of loops or procedures.
In the simple form, the term of the attribute must be
a term of SMT-LIB \terminal{Int} sort,
Whenever the statement is executed, the value of the term
must decrease strictly according to the built-in order~\terminal{<} but remain non-zero.
We do not aim to support further primitive well-founded orders,
rather other orders should be mapped to integers explicitly,
for example by converting bit-vectors to their numeric values,
or by taking the cardinality of a finite set.

A second form with attribute \terminal{:decreases-lex}
is supported to denote compound well-founded orders
by lexicographic combination of integer terms.
The leftmost entry is the most significant one.
The order is defined by a (possibly-empty) prefix of the entries that remain unchanged,
followed by an entry that decreases strictly.

Whenever the ranking function does not decrease on some execution,
or falls outside the designated range of values such as non-negative integers,
a counterexample to this must be exported,
see~\cref{ssec:counterexample-side-conditions}.

\subsection{(Not-)Recurring Locations (\livenessProperty)}

We have two main liveness properties, 
\terminal{:recurring} and \terminal{:not-recurring}.
They are used to annotate tags,
denoting locations which must be visited
infinitely often (recurring) or only finitely many
times (not-recurring)
on all infinite paths.
For finite paths, both properties are considered 
to be trivially
satisfied, which will become clear
from the formalization of the semantics in a future
version of the format.
For recurring locations, we can decide
which location of the tag group to visit
in each iteration, i.e., not all of them need
to be visited infinitely often.
In particular, these liveness properties 
are implicitly satisfied by finite runs.

Guaranteed termination of a program can be encoded by 
annotating each loop and function head as \terminal{:not-recurring}.
An alternative formalization is to assume an 
implicit self-loop at the end
of the program which does nothing but has the 
implicit tag \terminal{exit}
that occurs only there, in which case termination 
is encoded by this tag being \terminal{:recurring}.
We claim that both approaches are equivalent, 
also with respect to validation.

In contrast to safety counterexamples, liveness counterexamples 
are oftentimes more involved, since they require
an argument that the system always makes progress.
This usually requires a combination of invariants and
ranking functions.
Counterexamples for liveness are discussed 
in~\cref{ssec:counterexample-liveness}.

\section{\Language Witnesses}
\label{sec:witnesses}

A witness is a sequence of commands,
which can be combined with the original script
to generate a new valid \Language~script.
This guarantees that witness validation is the same
task as verification, but aided by the witness,
making it possible to reuse verification tools
for validation.
We distinguish between correctness, and violation witnesses.
Correctness witnesses certify 
that a given property holds,
whereas violation witnesses certify 
that a given property is violated.

Since both correctness and violation witnesses
are comprised of several components,
we delimit the response of the verifier inside a pair of parentheses.
This allows clients to reliably detect the end of a multi-line response
without accidentally blocking on the communication channel
and it allows tools in general to detect incomplete witnesses stored in files.
We emphasize that witnesses are part of the \emph{response}
of the verifier to a script and as such they
are not valid as commands in a \Language{} script.
\begin{align*}
\nonterminal{witness}
    \Coloneqq \nonterminal{correctness-witness} \alt \nonterminal{violation-witness}
\end{align*}

The goal of a witness is to provide sufficient information
to make the validation task inherently simpler 
than the original verification task.
For correctness witnesses, this is achieved by providing
inductive invariants, modular contracts, and ranking functions.
These allow the direct generation of verification conditions
in first-order logic, which can be checked by an SMT solver.
For violation witnesses, this is achieved by providing
a concrete execution trace, whenever possible, that leads 
to a property violation.
This allows the validation to be done by concrete execution
instead of symbolic reasoning.
The only exception is when a liveness property is violated,
where a lasso-shaped execution is required to witness the violation,
which again can be checked by an SMT solver.

In addition to the information to replay the verification verdict,
a witness may also contain metadata, like the tool that generated it,
the time of generation, and other useful information for debugging
and tracking purposes. For this, we use the SMT-LIB
\terminal{set-info} command to set arbitrary attributes.
We recommend setting at least the \terminal{:producer} and,
\terminal{:input-files} attributes to indicate the tool
that generated the witness, and the input files
which composed the verification task for which the
witness was created.
\begin{align*}
\nonterminal{metadata}
    \Coloneqq \LP\terminal{set-info}~\nonterminal{attribute}\RP^\rstar
\end{align*}

This section presents mostly the syntax of witnesses,
their semantics is kept informal for now, but will be 
made precise in a future version
of this document.
\Cref{ssec:correctness-witnesses} describes correctness witnesses,
\cref{ssec:counterexample-safety} describes counterexamples to safety,
and~\cref{ssec:counterexample-side-conditions} describes counterexamples 
to modular specifications, like inductiveness of invariants, modularity of contracts,
and the decreasing property of ranking functions, and
finally \cref{ssec:counterexample-liveness} describes counterexamples to liveness.

Whenever a violation witness is validated, it contains a trace
which needs to be fulfilled. This requires the trace to
not only be syntactically valid w.r.t. the program,
but also terminate, and reach the violation.
For correctness witnesses, the annotations provided
could contain unsound information like $\LP\terminal{assert}~\terminal{true}\RP$.
Invalid witnesses are discussed in~\cref{ssec:invalid-witnesses},
for both correctness, and violation witnesses.
For most of these cases, exporting a witness showing the
invalidity of the program composed of the original script and the witness
is left for future work.
Therefore, it is not possible to export a witness for all scripts describing
exactly what went wrong during validation.
A special case, where the witness can no longer be followed,
since some step through the program cannot be realized,
e.g., choosing a value outside the allowed range of a \terminal{choice} statement, 
is discussed in~\cref{ssec:path-non-existence-witness}.

\subsection{Correctness Witnesses}
\label{ssec:correctness-witnesses}

For both safety and liveness properties, 
a correctness witness is a sequence of 
$\nonterminal{smt-lib-command}$s
followed by a sequence of
$\terminal{annotate-tag}$
commands, as shown in~\cref{fig:correctness-witnesses}.
The SMT-LIB commands can be used to provide
auxiliary definitions, e.g., abbreviations for predicates used in invariants,
frame conditions,
or theory extensions, e.g., to express summations over array ranges.
These may be needed to express facts mathematical facts to
be used in the annotations proving the program correct.
Of the commands shown in \cref{fig:smt-lib-commands},
\Language accepts all those except for those starting with
\terminal{get-} and \terminal{set-}.

\begin{figure*}[t]
    \centering
    \centering
    \begin{minipage}{0.5\textwidth}
        \begin{align*}
            \nonterminal{correctness-witness}
                & \Coloneqq \LP \nonterminal{metadata}\rstar \\
                & \altspace \nonterminal{smt-lib-command}^\rstar~ \\
                & \altspace ~ \LP \terminal{annotate-tag}\ \tau\ a_1, \dots, a_n \RP^\rstar \RP
        \end{align*}
    \end{minipage}
    \caption{Syntax of correctness witnesses}
    \label{fig:correctness-witnesses}
\end{figure*}

We point out that the \terminal{assert} command is to be considered ``dangerous'',
insofar that only \emph{conservative} extensions to the theory
should be allowed here, e.g., postulating \terminal{(assert false)}
reduce the potential models to the empty set and thus vacuously
make the specification true.
We leave this problem for the future and suggest for the use of \Language
in competitions to simply forbid the \terminal{assert} command here.
However, note that SMT-LIB has other ways to introduce inconsistency into the specification,
for example by non-terminating functions declared with \terminal{define-fun-rec}.

Afterwards, the sequence of \terminal{annotate-tag} commands,
provides a modular abstraction for loops via invariants,
procedures via contracts,
and ranking functions for liveness properties.

To be able to annotate the required program locations,
the program is required to have tags at
\begin{enumerate*}[label=(\alph*)]
    \item the top-level statement in each procedure,
    \item each loop head, and
    \item each label
\end{enumerate*}.
These conditions can be checked by a linter, as discussed in~\cref{ssec:fully-well-formed-programs}.

\subsection{Safety Violations}
\label{ssec:counterexample-safety}

A violation of a safety property is witnessed by a collection of concrete counterexample traces.
Each trace indications one way to violate some property of the specification.
The grammar is shown in~\cref{fig:counterexample-syntax}.
Each trace is a sequence of entries that describe the initial conditions
at the falsified \terminal{verify-call}
followed by a detailed description of how to resolve the non-determism towards
the indicated property violation.

\begin{figure*}[t]
    \centering
    \begin{minipage}{0.0\textwidth}
        \begin{align*}
            \nonterminal{violation-witness}
                & \Coloneqq~ \LP \nonterminal{metadata}~\LP \terminal{select-trace}~\nonterminal{trace} \RP \rstar \RP \\[4pt]
            \nonterminal{trace}
                & \Coloneqq~
                        \LP \terminal{model}~\nonterminal{model-response}\rstar \RP \\
                & \altspace \LP \terminal{init-global-vars}~ \LP\nonterminal{symbol}~\nonterminal{value}\RP\rstar\RP \\
                & \altspace \LP \terminal{entry-proc}~   \nonterminal{symbol} \RP \\
                & \altspace     \LP \terminal{steps}~\nonterminal{step}\rstar\RP \\
                & \altspace \nonterminal{violated-property} \\
                & \altspace \LP \terminal{using-annotation}~\nonterminal{symbol}~
                                    \nonterminal{attribute}\rplus \RP\rstar \\[4pt]
            \nonterminal{step}
                & \Coloneqq~\LP \terminal{init-proc-vars}~ \nonterminal{symbol}\ \LP\nonterminal{symbol}\ \nonterminal{value} \RP\rstar \RP \\
                & \alt     ~\LP \terminal{havoc}~  \LP\nonterminal{symbol}\ \nonterminal{value} \RP\rstar \RP \\
                & \alt     ~\LP \terminal{choice}~  k \RP \\
                & \alt     ~\LP \terminal{leap}~\nonterminal{symbol}~\LP
                                    \nonterminal{symbol}~\nonterminal{value} \RP\rstar \RP \\[4pt]
            \nonterminal{violated-property}
                & \Coloneqq~\LP \terminal{incorrect-annotation}~
                                    \nonterminal{symbol}\
                                    \nonterminal{attribute}\rplus \RP \\
                & \alt     ~\LP \terminal{invalid-step}~\nonterminal{step} \RP
        \end{align*}
    \end{minipage}
    \caption{Syntax of concrete traces for counterexamples}
    \label{fig:counterexample-syntax}
\end{figure*}

Specifically, the first part of a trace is a trace-specific SMT-LIB model,
comprised of several SMT-LIB definitions (see $\nonterminal{model-response}$ in \cite[Sec. 3.11.1]{SMTLIB27}),
covering notably all global constant symbols.
The intention of including the model is to allow validation by concrete execution.
This model is paired with an initialization of some of global variables by name,
where $\LP \terminal{init-global-vars}\ \LP x_1\ v_1 \RP~\cdots~\LP x_n\ v_n\RP \RP$
can leave out those global variables whose initial value is irrelevant for
the concrete execution, like those global variables which are guaranteed to be
assigned or havoced explicitly before being read, on that execution path.

For good measure, the trace indicates the entry-point into the program in terms
of the procedure name of the corresponding \terminal{verify-call}.
We emphasize that listing the concrete values of input parameters here is not necessary,
as we have full information to determine these from the model and global variables already.
This idea is applied to representation of subsequent execution steps, defined by $\nonterminal{step}$,
insofar that only additional information is listed that is required to resolve
any type of (relevant) non-determinism.
The listed $\nonterminal{step}$s are thus matched eagerly to the statements occuring in the execution trace as followed.

A $\LP \terminal{init-proc-vars}\ \rho\ \LP x_1\ v_1 \RP~\cdots~\LP x_n\ v_n\RP \RP$ step
is applied right after a call to procedure with name~$\rho$,
initializing some of the local variables as well as output variables.
Similarly to the \terminal{init-global-vars} trace entry, we may omit those variables
whose initial value is irrelevant.
The rationale is that often, local variables and outputs are written explicitly,
thereby avoiding the need that a verifier invents arbitrary initial values in such cases,
just to have these overwritten immediately.
Note, we including the procedure name~$\rho$ for debugging purposes, it should be recoverable during execution, and the trace should be rejected if a call with a mismatching name is encountered.

The same idea is applied to \terminal{havoc} step, which are matched by \terminal{havoc}
statments but are presented here rather as an assignment to those variables for which
the initialization matters.
A \terminal{choice} step selects the branch leading to the violation from a \terminal{choice} statement by index (starting at 0).

A violation is claimed by trace elements
$\LP \terminal{incorrect-annotation}\ \tau\ a_1 \dots a_n \RP$
where $\tau$ is a tag uniquely identifying a position in the program
and $a_1 \dots a_n$ should all be violated by this trace.
Checking whether that annotation occurs syntactically exactly
as part of the original annotation
is an aspect of witness validation.
It may be possible to relax this check in the future
and to permit some reasoning, for example to allow signalling
a violation of $\phi$ for the specification
    $\terminal{:check-true}\ \LP \code{and}\ \phi\ \psi\RP$,
as the latter is semantically equivalent to the annotations
    $\terminal{:check-true}\ \phi\ \terminal{:check-true}\ \psi$.

Steps \terminal{leap} are only allowed to occur in counterexamples to inductiveness,
they will be described in \cref{ssec:counterexample-side-conditions}.
Similarly, \terminal{using-annotation} at the end of a trace is only allowed to occur
in counterexamples to liveness as explained in \cref{ssec:counterexample-liveness}.
Finally, \terminal{invalid-step}s refers to previously-claimed violation traces
that were found to be spurious as explained in \cref{ssec:invalid-witnesses,ssec:path-non-existence-witness}.

Confirming that traces lead to a violation should by design be possible with concrete execution.
However, a bad witness could deliberately lead such an approach into an infinite loop.
This problem is left for the future, e.g., by allowing for some approximation
of loop and recursion counters as part of the trace.
Another future extension to \Language would be a symbolic alternative to
describe counterexample traces, for example, that describe the program's unrolling
towards the violation.

\subsection{Violations of Modular Specifications}
\label{ssec:counterexample-side-conditions}

In contrast to counterexamples for \terminal{:check-true} annotations,
\terminal{:invariant}s and statement contracts are intended not just to
express true properties of all executions but rather admit an inductive proof.
This is a stronger property that is required to hold potentially on executions
that do not exists in reality but which are implicitly included 
from the overapproximation inherent to such constructs.
The same is true for ranking functions,
which are supposed to decrease in each iteration.

However, one should not be forced to 
include as part of invariants those properties
which are obviously stable across loop iterations,
notably when they are over variables not modified by the loop.
This leads to a somewhat subtle notion of 
inductiveness with respect to the given context.
As explained more detail in~\cref{ssec:invariants}.

As a consequence, counterexamples to 
inductiveness, and the decreasing property of 
ranking functions cannot just be execution traces.
Instead, one needs to be able to
over-approximate the effect of loops
and statement contracts in a way that is consistent
with the given annotations.
Through this, one ends up in a state 
from which a violation of the respective property can be observed.
This is achieved by the step inside a trace
$\LP \terminal{leap}~\nonterminal{symbol}~\LP \nonterminal{symbol}\ \nonterminal{value} \RP\rstar \RP$ 
which once the provided tag is reached, transitions into an arbitrary state at that tag,
where the values of the given symbols are set to the given values.
This allows the counterexample to go
into a hypothetical state that may not be reachable
by a concrete execution but which is consistent
with the over-approximation inherent to loops and statement contracts.
This requires the state given by the \terminal{leap} step
to satisfy all annotations at the given tag,
notably the invariants at loop heads,
and the pre-conditions of statement contracts.
After executing a \terminal{leap} step,
the execution continues as normal.
The \terminal{leap} command must be used
as soon as the corresponding tag is reached,
and before executing any statement at that tag.

The grammar of counterexamples to modular specifications,
is given in~\cref{fig:counterexample-syntax},
but in this case the \terminal{leap} step
is allowed as a step inside the trace,
to over-approximate the effect of loops and contracts.
Nonetheless, \terminal{using-annotation} constructs
are still not allowed here, and
the violated property must still always be \terminal{incorrect-annotation}.

Note that the \terminal{leap} step
should contain all, and only those variables
which are modified by the loop or statement
inside the contract.
According to our understanding of inductiveness,
all of these variables may take arbitrary values.

\subsection{Liveness Violations}
\label{ssec:counterexample-liveness}

A violation of a liveness property is witnessed by a lasso,
which has a concrete trace as a stem, just as for counterexamples to safety,
that navigates to the head of a loop that, if it has infinite executions,
fails to validate the property.
Their syntax is shown in~\cref{fig:counterexample-syntax},
where all the constructs in the grammar are allowed,
including \terminal{leap} and \terminal{using-annotation},
and the violated property must always be \terminal{incorrect-annotation}.

The stem of the lasso is given by a
$\LP \terminal{select-trace}~\nonterminal{trace} \RP$ command,
which navigates to the tag which corresponds
to the head of the loop resulting
in the lasso which witnesses the violation 
of the liveness property.
For the stem to be valid,
the trace must result in a state satisfying
all annotations set by the \terminal{using-annotation} 
elements for the tag at the end of the trace.
In case no such tag exists,
\terminal{true} is assumed.

The lasso itself is described by a sequence of
$\LP \terminal{using-annotation}~\dots\RP$ elements.
These elements describe how to reach
the head of the loop again,
by using the annotations to inductively abstract everything in between,
and using ranking functions to ensure progress
towards visiting the head of the lasso.
For the liveness property to be violated,
all paths starting in an arbitrary state at the head of the lasso
fulfilling all annotations set by 
the \terminal{using-annotation} elements there,
must always eventually return to the head of the lasso
by abstracting everything in between
using the annotations,
fulfilling all annotations there again.

Decoupling the original annotations from those
needed to witness a liveness violation is crucial
to allow for the verifier to detect liveness bugs
even if the original safety annotations are incorrect, too.
In particular, this requires the solver to ignore already
existing annotations, and to only consider those
set by the \terminal{using-annotation} elements
inside the \terminal{select-trace} command.

\subsection{Invalid Witnesses}
\label{ssec:invalid-witnesses}

Buggy verifiers could produce witnesses that do not make much sense,
for example, when annotations do not refer to existing locations,
which is easy enough to detect by syntactic checks.
Moreover, all correctness witnesses only extend the set of annotations,
which means that a problematic witness can be identified and flagged
by the standard means of verification,
and a counterexample to a supposed correctness witness can be represented and re-checked.
However, we currently do not have a certifiable way to refute that
counterexample traces exist in the program.
The problem is on one hand more subtle (lack of monotony when extending the verification task)
but on the other hand simpler, because detecting the non-existence of traces
requires simple techniques in theory (i.e., concrete execution).
To showcase how such cases could be handled, 
the case of not being able to execute
a $\nonterminal{step}$ in a trace is 
discussed in~\cref{ssec:path-non-existence-witness}.

We therefore reserve the standard response \terminal{error(cannot export witness)} to denote
invalid programs where the reason for the invalidity cannot be expressed
using \Language.
Verifiers are encouraged to include a human-readable error message.

\subsection{Violations to Step Execution}
\label{ssec:path-non-existence-witness}

Whenever a \Language~script contains a \terminal{select-trace}
command, it is possible that the trace cannot be executed
because some step cannot be fulfilled.
For example, a \terminal{choice} step
selects a value outside the allowed range.
Some of these constraints cannot be checked syntactically,
but require some semantic reasoning, by executing the program
along the trace.
When such a situation occurs, the verifier should return \terminal{incorrect},
and export a violation witness
that indicates that the trace cannot be executed.
The violation witness should correspond to a safety violation 
witness, see~\cref{ssec:counterexample-safety},
where the violated property at the end of the trace
is \terminal{invalid-step} indicating the step that could not be executed.

\section{Well-Formed \Language Programs}
\label{sec:well-formed-programs}

The syntax presented in this document defines 
a large set of possible \Language~programs.
However, not all syntactically valid programs are
appropriate as a verification task.
For example, a verifier may want to export an invariant,
but have no tag to attach it to, leaving it unable to export
a correctness witness.
Therefore, we define a set of conditions
that a \Language~program should satisfy
in order to be well-formed.
These rules will also be implemented in a linter
that can be used to check \Language~programs automatically.

The basis for \Language~programs are S-expressions as defined in \cite[Sects.~3.1--3.4]{SMTLIB27}.
For \Language~scripts, we distinguish between two levels of \emph{well-formedness} conditions.
Structural and typing conditions (\cref{ssec:basic-well-formed-programs})
which ensure that the program is executable,
and conditions for \emph{full well-formedness} (\cref{ssec:fully-well-formed-programs})
which in addition ensure that all annotations can be meaningfully interpreted
by a verifier or validator, ensuring that the program represents
a meaningful verification task.

\subsection{Structural Conditions and Typing}
\label{ssec:basic-well-formed-programs}

The conditions outlined in this section aim to ensure that each term, statement, and command
makes sense semantically and thus guarantee that the program to be executed.

Here, we deliberately exclude checking any property specifications
annotated inline in the program or using the top-level \terminal{annotate-tag} command.
The first (minor) difficulty with the latter is that the terms don't occur
within their respective lexical scopes.
More importantly, however, these terms can contain syntactic constructs beyond those in SMT-LIB,
specifically references to certain execution states via \terminal{at}, %
which may need some further context to be resolved or encoded and that context may be specific
to the verification tool or methodology.

All of these conditions should be checked by a compliant parser during the parsing process.
The selection is such that this is relatively straight-forward, both conceptually
as well as from an implementation view.

\begin{enumerate}[label=(\alph*)]
    \item it is built according to the syntax defined in this document,
    \item all requirements inherited from SMT-LIB are respected,
    \item all functions and procedures are declared, and/or defined before they are used,
        except for groups of explicit mutual recursion like those marked with
        \terminal{define-procs-rec},
    \item no symbol int he script, i.e., variable, function, tags, or procedure name,
        starts with the reserved symbol prefix \code{\#},
    \item all terms representing a condition of a statement, like
        \terminal{if}, \terminal{assume}, and \terminal{while} 
        must be of the boolean sort,
    \item the same procedure name is not used 
            for two \terminal{define-proc} commands,
    \item labels of \terminal{label} statements need to be uninque within a procedure
          (but not necessarily globally unique),
    \item labels of \terminal{goto} statements must exist within the enclosing procedure,
    \item all variables used in terms, statements, and top-level commands are in scope,
    \item all terms, statements, and top-level commands are type-correct, 
    \item all procedure calls provide the correct number of arguments,
    \item variables occurring as left-hand sides of assignments or recipients 
        of procedure returns are allowed to be modified: global, outputs, and local variables,
        of the enclosing procedure, but \emph{not} procedure inputs
\end{enumerate}

\subsection{Fully Well-formed \Language{} Programs}
\label{ssec:fully-well-formed-programs}

The additional constraints outlined below are concerned with characterizing
``useful'' verification tasks.
In particular, these conditions guarantee that all annotations can be
meaningfully interpreted by a verifier or validator.
The requirements on the presence of certain tags ensures that
important program locations can be referenced by witnesses.

\begin{enumerate}[label=(\alph*)]
    \item the program is structurally valid and type correct as defined in \cref{ssec:basic-well-formed-programs}
    \item program variables occurring in annotations must be in scope of all locations in the program they refer to, via the \terminal{:tag}-based linking mechanism
    \item function symbols occurring in inline statement annotations 
        must have been declared before the respective statement
    \item function symbols occurring in \terminal{annotate-tag} commands must be declared before that command, regardless of where the respective tag may occur (see \cref{ssec:annotate-tag}).
    \item all terms for the property annotations \terminal{:invariant}, \terminal{:check-true},
        \terminal{:requires}, and \terminal{:ensures} must be type-correct and of the boolean sort,
    \item all terms in a \terminal{:decreases} and \terminal{:decreases-lex}
          annotation must be type-correct and of the integer sort,
    \item all variables used in annotations are in scope at the respective tag,
    \item $\terminal{at}$ references can be statically resolved,
    \item the top-level statement in every procedure, each 
        loop, and every label has a script-unique tag, and
    \item each statement with an inline annotation 
        has at least one script-unique tag.
\end{enumerate}

\subsection{Warnings and Code Smells}

Furthermore, there are some \Language~scripts
which while well-formed present common pitfalls,
which may indicate a problem in the tool creating
or processing the \Language~scripts.
In particular, linters are encouraged to warn in case:
\begin{enumerate}[label=(\alph*)]
    \item some local variable is read before it has been explicitly initialized or havoced,
    \item some output variables are not initialized at procedure return,
    \item the version of the format is missing using a \terminal{set-info} command,
    \item the export of witnesses is not explicitlyy enabled or disabled using a \terminal{set-info} command.
\end{enumerate}

\section{Current Tools and Libraries}
\label{sec:tools}

One of the primary goals of \Language{} is to 
facilitate interoperability between different verification tools.
To this end, we provide some tools and libraries
that support working with \Language{} programs.
We envision that these can be used and extended 
by the community to build a rich ecosystem around \Language{}
aiming to fulfill the vision presented in~\cref{fig:transformation}.
We recommend looking at the official website of \Language{}
\webpageurl{} for the latest information
on available tools and libraries.

\subsection{Python Library: \pysvlib}

For working with \Language{} programs in Python,
we provide the \pysvlib 
library\footnote{\url{https://gitlab.com/sosy-lab/benchmarking/sv-lib/-/blob/format-1.0/pysvlib}}
as a python package. This package provides
some common functionality for parsing, analyzing,
and manipulating \Language{} programs.
Currently it contains a parser
that translates \Language{} scripts into an
abstract syntax tree (AST).
In addition, it contains a printer which
serializes the AST back into a string.
Furthermore, it includes a verification
condition generator which can be used
to implement a validator for \Language{} programs
and witnesses.
We defer further details to the documentation
of the library itself.

\subsection{ANTLR Grammar}

While python is a popular programming language,
other languages are also used in the verification community.
To facilitate working with \Language{} programs
in other programming languages,
we provide an \texttt{ANTLR} grammar from which
a parser for most languages can be easily generated.
The \texttt{ANTLR} grammar
for \Language{} programs\footnote{\url{https://gitlab.com/sosy-lab/benchmarking/sv-lib/-/blob/format-1.0/grammar/SvLib.g4}}
is based on an existing SMT-LIB grammar\footnote{\url{https://github.com/antlr/grammars-v4/blob/5660ba571209e7c28c0e36c38414729e5b6db087/smtlibv2/SMTLIBv2.g4}},
which it extends by adding the additional commands, statements,
and properties defined in this document.

\subsection{SV-LIB in \cpachecker}

The software verification tool \cpachecker
\cite{CPAchecker-3.0-tutorial} is a mature software
verification framework which allows developers to
easily implement new analyses and verification techniques.
Furthermore, its architecture allows developers to
integrate new input languages while making it possible
to reuse existing analyses and components
with only little extra effort.
Therefore, we have integrated \Language{}
as an input language into \cpachecker to enable
development of further analyses on a mature code-base.

This integration includes a front-end
that translates \Language{} programs first
into an AST representation of the program, which is 
then translated into \cpachecker's internal CFA representation.
The integration also includes a translation to and from \Language{} terms into
SMT-LIB formulas using the \tool{JavaSMT}~\cite{JavaSMT} interface.
Finally, we include support for exporting
witnesses
by translating \cpachecker's analysis result in 
form of an abstract reachability graph (ARG) or a counterexample
into a \Language{} correctness or violation witness.
With these components we implemented support for \Language
in \cpachecker's predicate analysis.
Our implementation was integrated into the
release 4.2 of \cpachecker~\cite{CPAchecker-4.2}.

\section{Conclusion}

We introduce the intermediate language \Language as
a standard exchange format
for software-verification tasks.
The language covers 
commands~(\cref{sec:commands}), statements~(\cref{sec:statements}), 
properties~(\cref{sec:properties}), and witnesses~(\cref{sec:witnesses}).
The goal of \Language~is to be
an intermediate language for imperative programs
with the goal of serving as exchange format between verification
frontends and backends.
This allows to decouple
the support of more programming languages (and their features)
from the implementation of verification algorithms, working
on the mathematical structure of programs.
Furthermore, we aim to provide a standard format
for different communities, focusing on the automatic
software verification and deductive verification communities.
We hope that \Language~will be adopted by the community
and will foster the exchange of verification tasks
between different tools and communities.

\inlineheadingbf{Data-Availability Statement}
The latest information on \Language can be found
online at the official website \webpageurl. This includes 
an \texttt{ANTLR} grammar for \Language and a \texttt{pip}
package with useful functionality to work with \Language
scripts.

\inlineheadingbf{Funding Statement}
This project was funded by the Deutsche Forschungsgemeinschaft (DFG)
--- \href{http://gepris.dfg.de/gepris/projekt/378803395}{378803395} (ConVeY)
and the Free State of Bavaria.

\inlineheadingbf{Acknowledgments}
\Language~was designed, and this article was written,
as direct response to a proposal made during a recent Dagstuhl seminar~\cite{Dagstuhl25-InfoEx},
in which a group of participants has proposed to design an exchange format like \Language.
We thank the community for the ideas, requirements, and support towards this proposal.
We are grateful for the inspiration that we received from K2~\cite{Kratos2}.
The development name for our language was~K3, before we finalized it to \Language.

\bibliographystyle{ACM-Reference-Format}
\bibliography{bib/dbeyer,bib/sw,bib/artifacts,bib/svcomp,bib/svcomp-artifacts,bib/testcomp,bib/testcomp-artifacts,bib/websites}

\end{document}